\documentclass[twocolumn]{aastex631}
\usepackage{natbib}
\usepackage{amsmath}
\usepackage[T1]{fontenc}
\usepackage[utf8]{inputenc}

\def\h{\hskip -0.3 mm}

\def\apjl{{ApJL}}
\def\apjs{{ApJS}}

\def\e#1{$\times$10$^{#1}$}

\def\lax{{$\mathrel{\hbox{\rlap{\hbox{\lower4pt\hbox{$\sim$}}}\hbox{$<$}}}$}}
\def\gax{{$\mathrel{\hbox{\rlap{\hbox{\lower4pt\hbox{$\sim$}}}\hbox{$>$}}}$}}
\def\simlt{\lower.5ex\hbox{$\; \buildrel < \over \sim \;$}}
\def\simgt{\lower.5ex\hbox{$\; \buildrel > \over \sim \;$}}

\def\cm2{cm$^{-2}$}

\def\l5100{$L_{5100}$}
\def\-->{$\rightarrow$}

\shorttitle{$M_*/L$ in NIR}
\shortauthors{Kim et al.}

\begin{document}

\title{Accuracy of Stellar Mass-to-light Ratios of Nearby Galaxies in the Near-Infrared}

\author[0000-0002-5857-5136]{Taehyun Kim}
\affiliation{Department of Astronomy and Atmospheric Sciences,
Kyungpook National University, Daegu 41566, Republic of Korea}

\author[0000-0002-3560-0781]{Minjin Kim}
\affiliation{Department of Astronomy and Atmospheric Sciences, 
Kyungpook National University, Daegu 41566, Republic of Korea}

\author[0000-0001-6947-5846]{Luis C. Ho}
\affiliation{Kavli Institute for Astronomy and Astrophysics, Peking University, Beijing 100871, People's Republic of China}
\affiliation{Department of Astronomy, School of Physics, Peking University, Beijing 100871, People's Republic of China}

\author[0000-0002-3309-8433]{Yang A. Li}
\affiliation{Department of Astronomy, School of Physics and Astronomy, Shanghai Jiao Tong University, Shanghai 200240, People's Republic of China}

\author[0000-0002-2770-808X]{Woong-Seob Jeong}
\affiliation{Korea Astronomy and Space Science Institute, Daejeon 34055, Republic of Korea}
\affiliation{University of Science and Technology, Korea, Daejeon 34113, Republic of Korea}

\author[0000-0002-6925-4821]{Dohyeong Kim}
\affiliation{Department of Earth Sciences, Pusan National University, Busan 46241, Republic of Korea}

\author[0000-0003-1647-3286]{Yongjung Kim}
\affiliation{Korea Astronomy and Space Science Institute, Daejeon 34055, Republic of Korea}

\author[0000-0003-1954-5046]{Bomee Lee}
\affiliation{Korea Astronomy and Space Science Institute, Daejeon 34055, Republic of Korea}

\author{Dongseob Lee}
\affiliation{Department of Earth Science Education, Kyungpook National University, Daegu 41566, Republic of Korea}

\author[0000-0003-3301-759X]{Jeong Hwan Lee}
\affiliation{Department of Astronomy and Atmospheric Sciences, 
Kyungpook National University, Daegu 41566, Republic of Korea}
\affil{Department of Physics and Astronomy, Seoul National University, 1 Gwanak-ro, Gwanak-gu, Seoul 08826, Republic of Korea}
\affil{Research Institute of Basic Sciences, Seoul National University, Seoul, 08826, Republic of Korea}

\author{Jeonghyun Pyo}
\affiliation{Korea Astronomy and Space Science Institute, Daejeon 34055, Republic of Korea}

\author[0000-0002-4179-2628]{Hyunjin Shim}
\affiliation{Department of Earth Science Education, Kyungpook National University, Daegu 41566, Republic of Korea}

\author[0000-0002-5346-0567]{Suyeon Son}
\affiliation{Department of Astronomy and Atmospheric Sciences, 
Kyungpook National University, Daegu 41566, Republic of Korea}

\author[0000-0002-4362-4070]{Hyunmi Song}
\affiliation{Department of Astronomy and Space Science, Chungnam National University, Daejeon 34134, Republic of Korea}

\author[0000-0003-3078-2763]{Yujin Yang}
\affiliation{Korea Astronomy and Space Science Institute, Daejeon 34055, Republic of Korea}
\affiliation{University of Science and Technology, Korea, Daejeon 34113, Republic of Korea}

\correspondingauthor{Minjin Kim}
\email{mkim.astro@gmail.com}

\begin{abstract}
Future satellite missions are expected to perform all-sky surveys, thus providing the entire sky near-infrared spectral data and consequently opening a new window to investigate the evolution of galaxies. Specifically, the infrared spectral data facilitate the precise estimation of stellar masses of numerous low-redshift galaxies. We utilize the synthetic spectral energy distribution (SED) of 2853 nearby galaxies drawn from the DustPedia (435) and Stripe 82 regions (2418). The stellar mass-to-light ratio ($M_*/L$) estimation accuracy over a wavelength range of $0.75-5.0$ $\mu$m is computed through the SED fitting of the multi-wavelength photometric dataset, which has not yet been intensively explored in previous studies. We find that the scatter in $M_*/L$ is significantly larger in the shorter and longer wavelength regimes due to the effect of the young stellar population and the dust contribution, respectively. While the scatter in $M_*/L$ approaches its minimum ($\sim0.10$ dex) at $\sim1.6$ $\mu$m, it remains sensitive to the adopted star formation history model. Furthermore, $M_*/L$ demonstrates weak and strong correlations with the stellar mass and the specific star formation rate (SFR), respectively. Upon adequately correcting the dependence of $M_*/L$ on the specific SFR, the scatter in the $M_*/L$ further reduces to $0.02$ dex at $\sim1.6$ $\mu$m. This indicates that the stellar mass can be estimated with an accuracy of $\sim0.02$ dex with a prior knowledge of SFR, which can be estimated using the infrared spectra obtained with future survey missions.          
\end{abstract}


\section{Introduction}
The stellar mass is one of the most fundamental parameters that offer crucial insights into the formation and evolution of galaxies. Furthermore, it helps probe the mass assembly history of galaxies through the stellar mass function \cite[e.g.,][]{madau2014}. In addition, its correlations with other physical properties of galaxies (e.g., metallicity, size, halo mass, and the mass of the central supermassive black hole) help facilitate the detailed evolution of galaxies \cite[e.g.,][]{maiolino2019,wechsler2018,kormendy2013,conselice2014}. Therefore, robust estimations of the stellar mass are crucial for understanding the detailed evolution of the galaxy.

\begin{figure}[tbp!]
\centering
\includegraphics[width=0.45\textwidth]{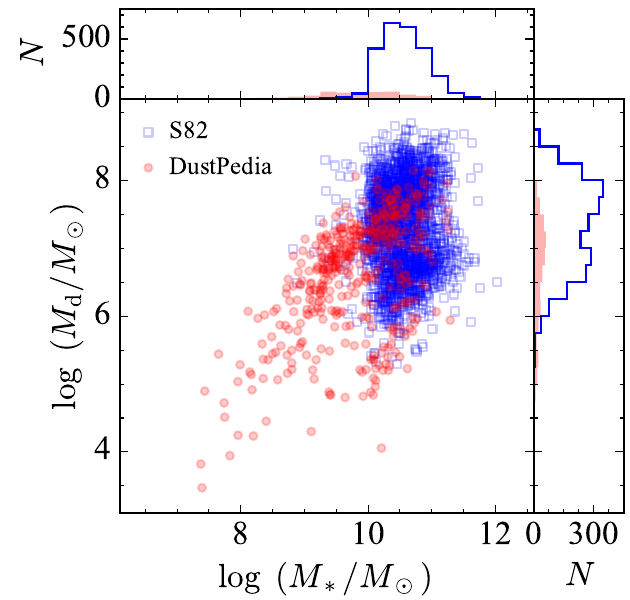}
\caption{
Distributions of stellar and dust masses derived from the SED fitting of multi-wavelength data spanning the UV to FIR. Red circles denote the sample from the DustPedia, and blue squares represent the sample from the SDSS S82 (\citealp{li2023}).}
\end{figure}

The stellar mass for a large number of samples has conventionally relied on broad-band photometry, assuming a constant mass-to-light ratio ($M_*/L$) for a given broad-band filter. In line with these practices, $M_*/L$, known to depend on the stellar age, metallicity, initial mass function (IMF), and dust extinction, was theoretically or semi-empirically derived using the stellar population model \cite[e.g.,][]{larson1978, bell2003, bruzual2003, zibetti2009, taylor2011}. As those properties cannot be directly measured with ease, a color index has been widely used as a secondary parameter to infer $M_*/L$. While the accuracy of this method strongly depends on the effective wavelength of the filter, the scatter in $M_*/L$ is $\sim0.1-0.2$ dex \cite[e.g.,][]{bell2003, zibetti2009,taylor2011,into2013,roediger2015}.  Note that this scatter does not account for systematics arising from uncertainties in the stellar population models, which can be up to $0.1-0.4$ dex (e.g., \citealp{conroy2013, hon2022}). This effect, specifically in NIR ($0.75-5\ \mu$m), is examined in another study (Lee et al., submitted). Alternatively, $M_*/L$ is often estimated using spectral energy distributions (SED) constructed from multiple broad-band photometric data (\citealp{conroy2013}). In this approach, synthetic SEDs are modeled with varying star formation history (SFH), which can introduce additional uncertainty in the stellar mass estimation. This uncertainty becomes more significant when photometric coverage is limited.        
\begin{figure}[tbp!]
\centering
\includegraphics[width=0.45\textwidth]{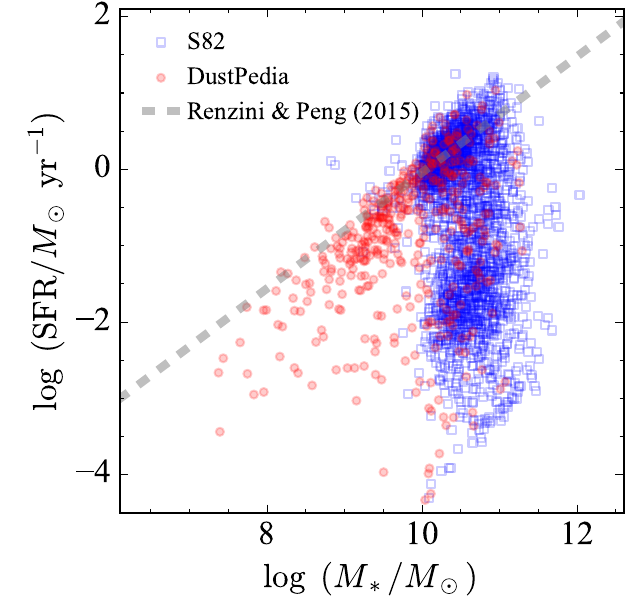}
\caption{
Distributions of stellar masses and SFRs derived from the SED fitting of multi-wavelength data spanning the UV to FIR. The symbols are the same as in Figure 1. The dashed line denotes the star-formation main sequence of nearby galaxies at $0.02 < z < 0.08$ (\citealp{renzini2015}).}
\end{figure}

\begin{figure*}[bpt]
\centering
\includegraphics[width=0.45\textwidth]{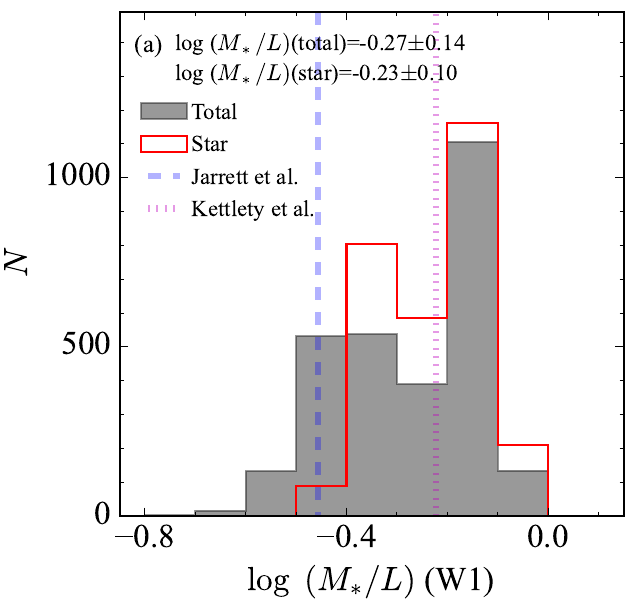}
\includegraphics[width=0.45\textwidth]{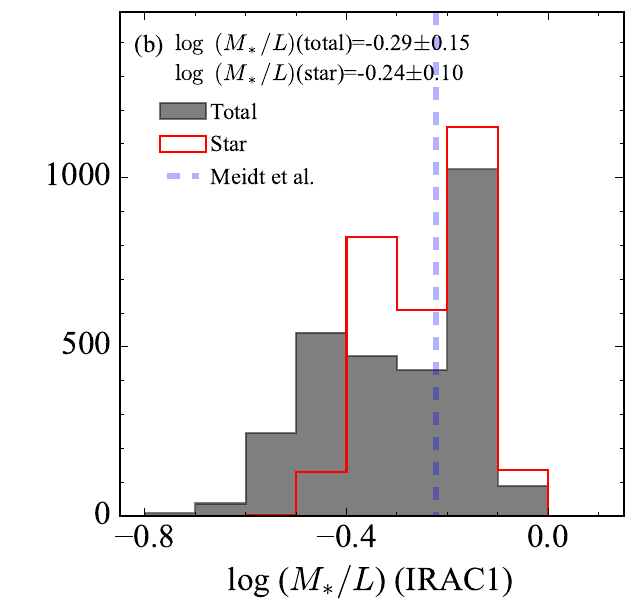}
\caption{
Distributions of $M_*/L$ ratios at the W1 (a) and IRAC1 (b) bands. The gray-shaded histograms represent $M_*/L$ derived from the total luminosity, encompassing stellar light, dust emissions, and nebular emissions. The red open histograms denote $M_*/L$ derived using the stellar luminosity. Mean and standard deviation values are shown in each panel.
(a) The red dotted and blue dashed lines indicate $M_*/L$ from \cite{kettlety2018} and \citet{jarrett2023}, respectively. (b) The blue dashed line denotes $M_*/L$ from \citet{meidt2014}.
}
\end{figure*}

Photometric data from the near-infrared (NIR) spectral regions are particularly useful to robustly estimate the stellar mass, as it is minimally affected by the dust attenuation and the $M_*/L$ ratio is less sensitive to the stellar age compared to those from the UV/optical data (\citealp{into2013}). However, intermediate-age stars such as those on the thermally pulsating asymptotic giant branch (TP-AGB) can significantly contribute to the fluxes in the NIR \cite[e.g.,][]{meidt2012},  underscoring that the accuracy of the stellar mass estimation is closely tied to the reliability of the stellar population models in the NIR \cite[e.g.,][]{taylor2011,conroy2013}. For example, $K$-band photometric data combined with $M/L_K$ can be a powerful tool for the stellar mass estimation in the local Universe \cite[e.g.,][]{bell2003}. In addition, NIR photometric data predominantly extracted from the {\it Spitzer} Space Telescope (\citealp{werner2004}) and the WISE mission (\citealp{wright2010}) contributed significantly to the stellar mass estimation of relatively nearby galaxies \cite[e.g.,][]{meidt2014, querejeta2015, jarrett2023}. Notably, NIR-based $M_*/L$ above $3\mu$m are easily contaminated by the emission lines and continuum originating from the warm dust, which requires empirical corrections using the NIR colors (e.g., W1-W2 or IRAC1-IRAC2). However, the complex contributions from the 3.3 $\mu$m polycyclic aromatic hydrocarbon (PAH) emission to the W1 and IRAC1 bands complicate the dust-contamination corrections solely based on NIR colors \cite[e.g.,][]{lee2012, yamada2013, inami2018,lai2020}, potentially introducing systematic uncertainties on the stellar mass estimations. Collectively, these effects can limit the accuracy of the stellar mass estimations based on NIR broad-band photometric data.

Upcoming satellite missions, including The Spectro-Photometer for the History of the Universe, Epoch of Reionization and Ices Explorer (SPHEREx), are poised to enrich our understanding of the Universe
(\citealt{korngut2018,crill2020}). In particular, SPHEREx will conduct an all-sky survey with linear variable filters, providing the all-sky spectral data spanning a wavelength range of $0.75-5\ \mu$m with spectral resolutions of $\sim40-130$ (\citealp{dore2016, crill2020}). This extensive dataset will enable us to estimate the stellar mass of numerous galaxies, which is crucial to understanding the galaxy evolution in the nearby Universe. Furthermore, leveraging its wide wavelength coverage and spectral capabilities, this NIR dataset is expected to facilitate precise stellar-mass estimations. However, the NIR spectral range (1–5 µm) remains relatively underexplored owing to limited observational data \cite[][]{brown2014}.

Based on the above backgrounds, we investigate the $M_*/L$ within the spectral range of $0.75-5.0\ \mu$m, covered by SPHEREx, which will provide the low spectral resolution ($R\sim40$) data. For that purpose, we utilize the synthetic SED obtained from the SED fitting of the multi-wavelength data of nearby galaxies adopted from literature. The sample selection and adopted dataset are summarized in Section 2. The NIR-based $M_*/L$ estimations are calculated in Section 3. The estimation accuracy of the obtained $M_*/L$ ratio is described in Section 4. In Section 5, we summarize the conclusions of this study. Throughout the paper, we adopt the cosmological parameters: $H_0=100h=70$ km ${\rm s}^{-1}$ ${\rm Mpc}^{-1}$ and $\Omega_\Lambda=0.7$. All magnitudes are given in the Vega system.

\section{Sample and Data}
\subsection{Sample Selection}
To utilize the SED derived from the observed photometric dataset, we first employ the sample from the DustPedia (\citealp{davies2017,clark2018}). The multi-wavelength dataset of the DustPedia spanning the UV to far-infrared (FIR) enables robust estimations of both stellar population and dust properties through SED fitting with the Code Investigating GALaxy Emission (CIGALE) code (\citealp{boquien2019}). This characteristic makes the DustPedia sample ideal for our study (\citealp{nersesian2019}). The DustPedia constructed the SEDs based on the aperture-matched photometry on the 42 bands data covering UV to sub-mm, obtained from the Hershel observation, along with the archival observations from the GALEX, SDSS, 2MASS, WISE, {\it Spitzer}, and {\it Planck}. The original DustPedia sample contains 875 nearby galaxies with $D_L \leq 79$ Mpc. However, this DustPedia sample is slightly biased toward low-mass galaxies ($\langle M_* \rangle = 10^{9.7\pm0.95} M_\odot$) because of their proximity. %Furthermore, the active galactic nuclei (AGN) component was not properly considered in the SED fitting for the DustPedia sample.  

To compensate for this limitation, we additionally adopt a sample of nearby ($0.01 \leq z \leq 0.11$) massive galaxies from \citet{li2023}, which employed data from the SDSS Stripe 82 region (S82). \citet{li2023} conducted the sophisticated photometry on the multi-wavelength imaging data of 2685 massive galaxies ($10^{10} M_\odot \lesssim M_* \lesssim 10^{11.5}  M_\odot$). As the initial sample selection was based on stellar mass estimates from the GALEX–SDSS–WISE Legacy Catalog (GSWLC-2; \citealp{salim2018}), a small fraction of the S82 sample finally shows $M_* < 10^{10} M_\odot$. These data encompassed UV, optical, NIR, MIR, and FIR data obtained from GALEX, SDSS, 2MASS, WISE, and Herschel-SPIRE, respectively. The matched-aperture and profile-fitting photometry were applied for the shorter-wavelength bands ranging from UV to W2 and the longer-wavelength bands, respectively.  
In addition, multi-band imaging decomposition was carefully performed to accurately remove the contribution from neighboring galaxies or foreground stars. These procedures resulted in reliable flux measurements and precise estimations of the uncertainties of the multi-wavelength dataset. From the comparison with the photometric data of GALEX, SDSS, and 2MASS extended source catalogs (\citealp{jarrett2000}), \cite{li2023} showed that their measurements agree within $\pm0.1$ mag in the UV, optical, and NIR bands. However, the photometric measurements of \cite{li2023} are systematically brighter ($\sim 0.2-1.0$ mag) than those in 2MASS point source catalog (\citealp{cutri2003}) and ALLWISE, in which the photometric data were originally estimated from a PSF-fitting method. Notably, these systematic discrepancies are significantly reduced up to $\leq 0.2$ mag when compared with the unWISE catalog (\citealp{lang2016}).           

\begin{figure*}[t!]
\centering
\includegraphics[width=0.45\textwidth]{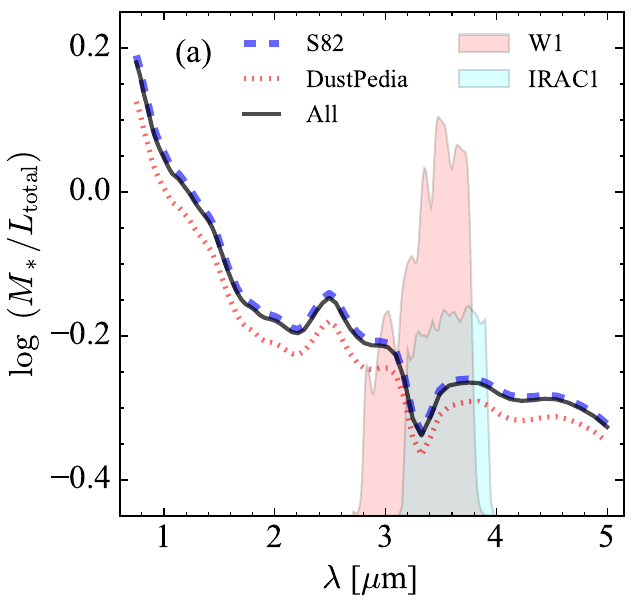}
\includegraphics[width=0.45\textwidth]{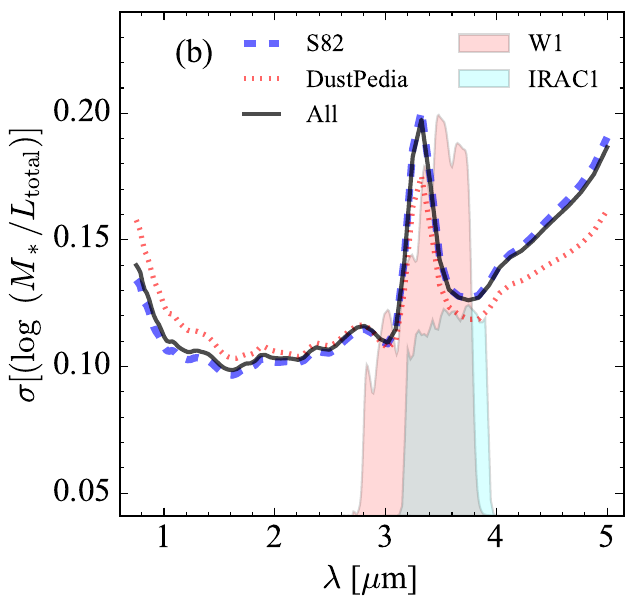}
\caption{
$M_*/L$ ratios calculated using the total luminosity (a) and their scatters (b) as functions of the wavelength. Subsamples from DustPedia and S82 are denoted by a red dotted line and a blue solid line, respectively. The thick solid line represents $M_*/L$ for the entire sample. The red and cyan areas denote transmission curves of W1 and IRAC1, respectively. Note that the unit of the transmission curves is arbitrary. Due to the 3.3 $\mu$m PAH emission, there appears to be a dip and a bump at 3.3 $\mu$m of the $M_*/L$ value and the scatter, respectively.}
\end{figure*}

\subsection{SED Fitting}
For the S82 sample, stellar and dust masses are drawn from \citet{li2023}. These parameters were obtained via SED fitting using the CIGALE. In this SED fitting process, the stellar SED was modeled using the simple stellar population (SSP) from \citet{bruzual2003}. The stellar population was modeled with a double-exponential SFH, comprising an old stellar population with an age of 12 Gyr and a young stellar population with an age range from 10 to 5000 Myr. 
The stellar emission was modeled with metallicities of $0.004-0.02$. The IMF of \citet{chabrier2003} was adopted. In addition, the starburst attenuation curve from \citet{calzetti2000} with a modification of the power-law slope was adopted to account for the dust attenuation in the SED fitting. 

Dust emission is modeled with $U_{\rm min}=0.1-25$, $q_{\rm PAH}=2.5$ (fixed), $\gamma=0.001-0.2$. Here, $U_{\rm min}$ denotes the minimum radiation field from the stars; $q_{\rm PAH}$ represents the mass fraction of PAHs; $\gamma$ indicates the mass fraction of dust illuminated from the minimum to maximum radiation field. $\alpha$, the power-law slope of the dust continuum over the IR wavelength, is fixed to 2.
Nebular emission lines were also included (\citealp{inoue2011}). The AGN component modeled with \citet{fritz2006} was incorporated. Here, the AGN fraction ($f_{\rm AGN}$) was calculated as the ratio of the AGN luminosity to the total IR luminosity estimated within $1-1000$ $\mu$m. 

Whereas the SED fitting with the CIGALE was also applied to the DustPedia sample (\citealp{jones2013, nersesian2019}), parameters for the stellar population and dust modeling significantly differ from those of \citet{li2023}. In addition, contrary to the S82 sample, the SED fitting of the DustPedia sample did not consider any AGN components. To avoid possible systematics due to these factors, we perform the SED fitting on the DustPedia dataset. To maintain consistency, we adopt the same fitting parameters and bandpasses used for the S82 dataset. We exclude the photometric data contaminated by artifacts or nearby sources, as well as data obtained from the imaging data that partially covers the target source (\citealp{clark2018}). To secure accurate stellar mass measurements from the SED fitting, samples lacking flux measurements in the optical band ($gri$) are further discarded. The systematic effects due to the discrepancy in the fitting parameters between the original DustPedia study (\citealp{nersesian2019}) and this study are further discussed in Appendix A.

We discard the targets with a reduced $\chi^2$ value exceeding 2 to ensure reliable assessments of the stellar and dust properties. Furthermore, following the recipe from \citet{li2023}, the AGNs defined as $f_{\rm AGN} > 0.1$ were excluded, as the flux contribution from the AGN can introduce additional uncertainties. This results in the final samples containing 435 and 2418 galaxies from the DustPedia and S82, respectively. Throughout this study, we utilize the synthetic SEDs obtained from the best fit for magnitude estimations across various filters and wavelengths. The goodness-of-fit, particularly in NIR regions, is examined in Appendix B. The stellar masses derived from the SED fitting agree with those of GALEX-SDSS-WISE Legacy Catalog 2 (GSWLC-2; \citealp{salim2018}) with $\Delta M_* = 0.01 \pm 0.11$ dex (\citealp{li2023}). However, a moderate difference is observed in the star formation rate when compared to GSWLC-2 ($\Delta \log\ {\rm SFR} = -0.12 \pm 0.56$ dex). As described in Li et al. (2023), this discrepancy arises from the differences in the methods of photometric measurements, particularly in the mid- and far-infrared bands.

\begin{figure*}[t!]
\centering
\includegraphics[width=0.45\textwidth]{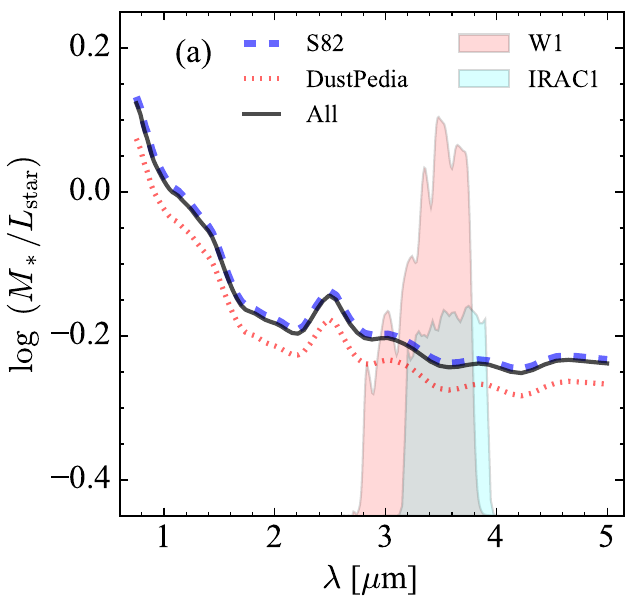}
\includegraphics[width=0.45\textwidth]{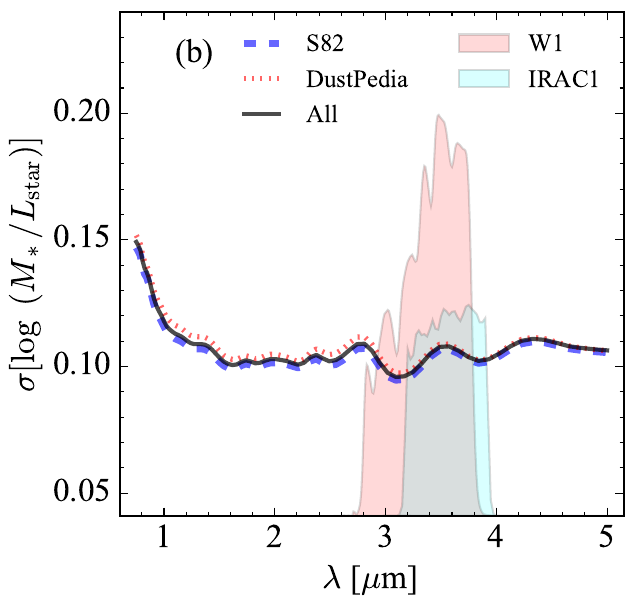}
\caption{
$M_*/L$ calculated using the stellar luminosity (a) and their scatters (b) as functions of the wavelength. The symbols are the same as in Figure 4.}
\end{figure*}

%Finally, the starburst attenuation curve from \citet{calzetti2000} with a modification of the power-law slope was adopted to account for the dust attenuation in the SED fitting. Nebular emission lines were also included (\citealp{inoue2011}). For the S82 sample, the AGN component modeled with \citet{fritz2006} was incorporated. Here, the AGN fraction ($f_{\rm AGN}$) was calculated as the ratio of the AGN luminosity to the total IR luminosity estimated within $1-1000$ $\mu$m. We discard the targets with a reduced $\chi^2$ value exceeding 2 to ensure reliable assessments of the stellar and dust properties.  Furthermore, the AGNs defined as $f_{\rm AGN} > 0$ were excluded, as the flux contribution from the AGN can introduce additional uncertainties. In contrast to the S82 sample, the SED fitting of the DustPedia sample did not consider any AGN components. To minimize AGN contamination, the objects from the DustPedia with $M_* \leq 10^{10} M_\odot$ are employed for further analysis. This mass threshold is imposed because the Eddington ratio of the AGN within this mass regime is significantly smaller than that of the AGN with massive supermassive black holes, resulting in negligible AGN contributions \cite[e.g.][]{she2017, bi2020, greene2020}.  

Figure 1 illustrates the distributions of stellar and dust masses determined from the SED fitting process, demonstrating the complementary characteristics of the two subsamples in the two-parameter space. Furthermore, our sample covers a wide range of SFR at a given stellar mass, from the star-forming main sequence to relatively quiescent galaxies, which is crucial to investigate the variations in the $M_*/L$ ratio with various properties of galaxies (Fig. 2).  

\section{Results}
\subsection{Mass-to-Light Ratio in Broadband Photometry}
Conventionally, stellar-mass estimations in the NIR band ($3-5\ \mu$m) have been based on broadband photometry obtained from the {\it Spitzer} Space Telescope and the WISE mission. As the first step of our analysis, we compute $M_*/L$ at commonly used broadband filters in the NIR (e.g., IRAC1 and W1).  Figure 3 shows $M_*/L$ at IRAC1 and W1 using the stellar and total luminosities. Overall, we can estimate the stellar mass with an accuracy of $0.14-0.15$ dex solely with the NIR data, which is consistent with the previous studies \cite[e.g.][]{meidt2014,kettlety2018, jarrett2023}.\footnote{It is worthwhile to note that this accuracy is calculated relative to the stellar mass derived from the SED fitting performed using the parameterized SFH, potentially introducing a zero-point offset of up to $0.3-0.4$ dex. This bias, however, is not considered in this study \cite[e.g.][]{conroy2013,lower2020}.} However, as the dust continuum and 3.3 $\mu$m PAH emission can contribute to the total luminosity in those filters, the $M_*/L$ ratios calculated using the total luminosity exhibit a substantially larger scatter compared to those calculated using the stellar luminosity. 

Meanwhile, considering only the stellar light, the scatter in $M_*/L$ can be significantly reduced down to $\sim0.10$ dex. While the average $M_*/L$ ratio at IRAC 1 [$M_*/L\sim0.58$ or $\log\ (M_*/L)\sim-0.24$] is in good agreement with the prediction for the old stellar population ($M_*/L \sim0.60$; \citealt{meidt2014, querejeta2015}), that at the W1 band [$M_*/L \sim0.54$ ($\log\ (M_*/L)\sim-0.27$) and $\sim0.59$ ($\log\ (M_*/L)\sim-0.23$] from the total and stellar luminosity, respectively] is significantly larger than the empirically derived value for the nearby galaxies ($M_*/L \sim0.35$; \citealp{jarrett2023}). Note that $M_*/L$ in \citet{jarrett2023} is derived from the W1 total luminosity. With the EAGLE simulations, \citet{norris2016} examined the M/L distribution of simulated early-type/quiescent galaxies and found $M_*/L \sim0.85$, which is in good agreement with a peak in the higher $M_*/L$ found in the distribution of $M_*/L$ (Fig. 3).   
While the origin of this discrepancy remains rather unclear, it can be partially attributed to differences in flux measurements across the chosen datasets. Specifically, the dedicated flux measurements of the WISE dataset obtained from \cite{li2023} are systematically brighter by $0.2-0.5$ mag compared to the ALLWISE measurements based on profile-fitting results, which was used to estimate the $M_*/L$ values in \citet{jarrett2023}. Interestingly, in our study, the $M_*/L$ values calculated for W1 and IRAC1 are consistent with each other, which supports the reliability of our estimates. It is also worthwhile noting that, unlike other studies, \citet{jarrett2023} used the observed total luminosity, instead of the stellar luminosity, to calculate the $M_*/L$ values, naturally resulting in a low $M_*/L$.

\begin{figure*}[tbp!]
\centering
\includegraphics[width=0.45\textwidth]{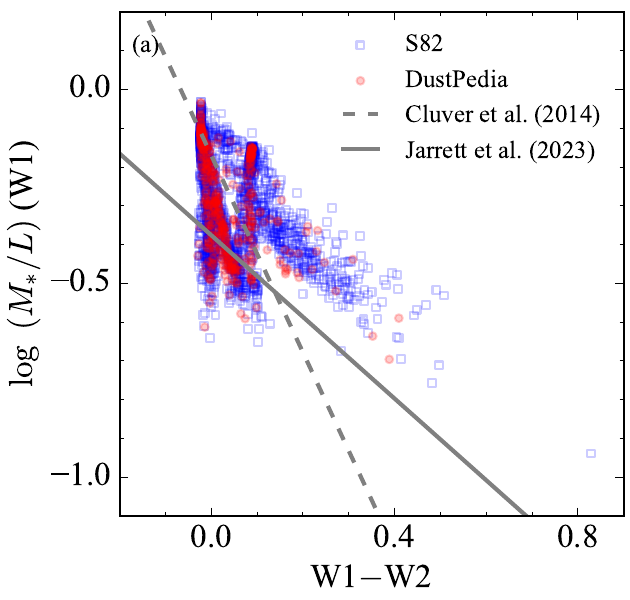}
\includegraphics[width=0.45\textwidth]{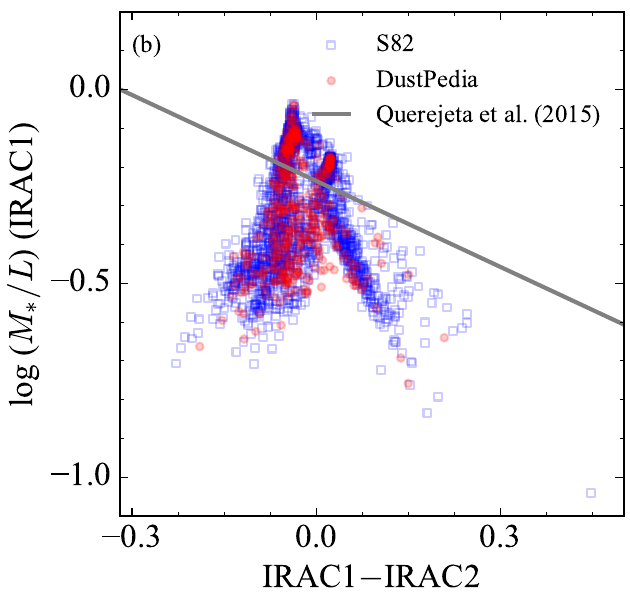}
\caption{
Dependence of $M_*/L$ calculated using the total luminosity on the W1$-$W2 color (a) and IRAC1-IRAC2 color (b). (a) The dashed line represents the empirical relation between $M_*/L$ and the W1$-$W2 color for $z\sim0.5$ galaxies from GALAXY AND MASS ASSEMBLY (GAMA) survey (\citealp{cluver2014}). The solid line denotes the newly calibrated relation for nearby galaxies at $z\sim0.15$ from the GAMA survey (\citealp{jarrett2023}). (b) The solid line represents the relation based on the sample from the Spitzer Survey of Stellar Structure in Galaxies (S$^4$G; \citealp{querejeta2015}).}
\end{figure*}

\subsection{Mass-to-Light Ratio in Spectral Data}
As the NIR spectral data will be obtained through future space missions, such as SPHEREx, investigating the $M_*/L$ in these spectral elements is worthwhile. To this end, we employ 68 spectral components with a spectral resolution of 40 (i.e., $\lambda/\Delta  \lambda \approx 40$, where $\Delta \lambda$ is the bandwidth of each spectral component), covering a wavelength range of 0.75$-$5.0 $\mu$m. For simplicity, the transmission curve of each spectral component is modeled using a Gaussian profile. Although this assumption does not exactly match the spectral channels of the SPHEREx, it is adequate for examining the overall trend of the $M_*/L$ ratio as a function of the wavelength. 

As expected, the $M_*/L$ ratio is inversely proportional to the wavelength because the $M_*/L$ ratios in the shorter wavelengths are more sensitive to the young stellar population compared to those at the longer wavelength. However, the scatter of the $M_*/L$ ratio exhibits a U-shaped pattern. The $M_*/L$ values calculated using the total luminosity are significantly affected at longer wavelengths due to the existence of the dust continuum and PAH emission (Fig. 4). In particular, the excess in the scatter is remarkable around 3.3 $\mu$m, highlighting the need for the careful removal of the PAH emission to facilitate accurate stellar mass estimations. This effect can be rigorously quantified using the spectral dataset provided by the SPHEREx mission, particularly for star-forming galaxies \cite[e.g.][]{xie2018, xie2018b, zhang2021}. 

\begin{figure}[tbp!]
\centering
\includegraphics[width=0.45\textwidth]{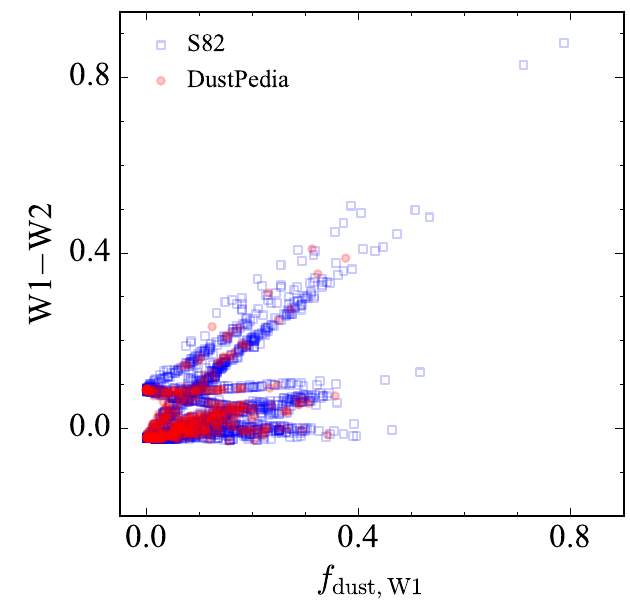}
\caption{
W1$-$W2 color as a function of the ratio of the dust flux to the total flux at the W1 band. The symbols are the same as in Figure 1.}
\end{figure}

%Intriguingly, the scatter in the $M_*/L$ ratios in the DustPedia sample are substantially greater than those in the S82 sample from \cite{li2023}. While the origin of this systematic discrepancy remains unclear, one possibility is that the adopted SFHs of low-mass galaxies, predominantly included within the DustPedia sample, are more complex compared to those of massive galaxies. Alternatively, the photometric data from the DustPedia sample can be less accurate than that from \cite{li2023}. This is because the aperture-based photometry of the DustPedia sample relied on a heterogeneous multi-wavelength imaging dataset, thus potentially introducing systematic uncertainties in the photometric measurements. For example, using the mean value of a sky annulus around the target galaxies as the sky background level for the DustPedia data could overestimate the sky level and thus underestimate the galaxy flux \cite[e.g.][]{li2011}. Meanwhile, for the S82 sample, the source-masked sky background was fitted with a 2D polynomial function for the optimal background subtraction. In addition to these differences, the methods of the photometric error estimation also differed across both subsamples. Specifically, while \citet{clark2018} considered the confusion noise as the sole source of photometric uncertainty in the DustPedia sample, \citet{li2023} considered additional sources such as the background, Poisson, and pixel-correlated noise (in the 2MASS and WISE datasets) for the S82 sample. This discrepancy in the error estimation techniques can also affect the SED fitting results.   
\begin{figure}[tbp!]
\centering
\includegraphics[width=0.45\textwidth]{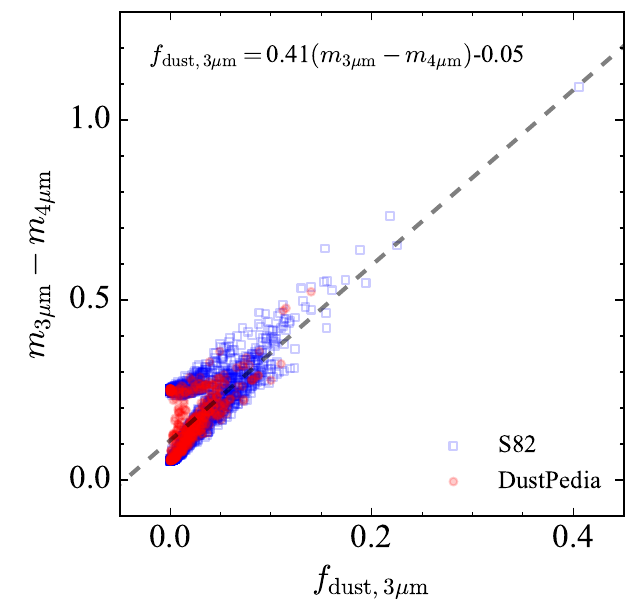}
\caption{
Correlation between the flux ratio the dust at $3\ \mu$m ($f_{\rm dust, 3\mu m}$) and the color index between 3 $\mu$m and 4 $\mu$m ($m_{\rm 3\mu m} - m_{\rm 4\mu m}$). The dashed line denotes the linear fit to the data.}
\end{figure}

\begin{figure}[tbp!]
\centering
\includegraphics[width=0.45\textwidth]{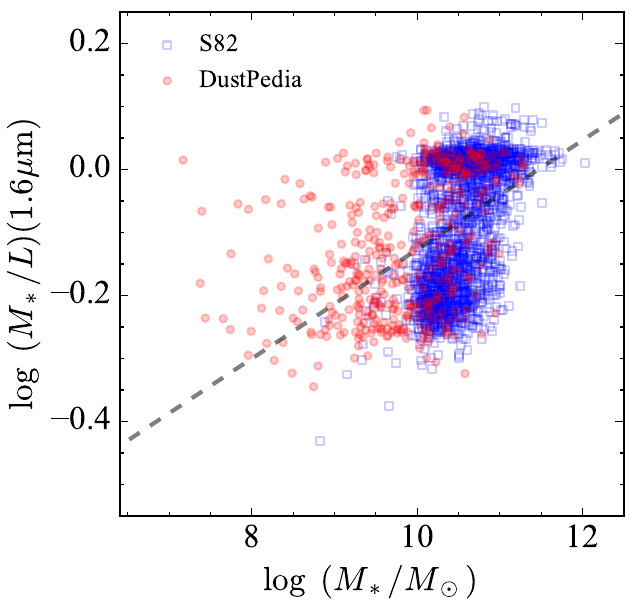}
\caption{
Correlation between $M_*/L$ and stellar mass ($M_*$) at $\sim1.6$ $\mu$m. Note that $M_*/L$ is calculated using the stellar luminosity. The dashed line represents the linear regression fit to the entire data.}
\end{figure}

In addition, the $M_*/L$ values in the DustPedia sample are systematically smaller than those in the S82 sample. 
%This can be attributed to two factors. Firstly, the $M_*/L$ is sensitive to the adopted IMF. Previous studies demonstrated that the Salpeter IMF overpredicts the $M_*/L$ values compared to the Charbier IMF \cite[e.g.,][]{madau2014}. However, this trend contradicts our finding (i.e., the SED fitting results for the DustPedia sample with the Salpeter exhibit smaller $M_*/L$ than those for the S82 sample with the Charbier IMF). 
The $M_*/L$ values are known to be strongly dependent on the specific SFR \cite[e.g.,][]{portinari2004}. The DustPedia sample tends to have a {\rm larger} specific SFR than the S82 sample (see \S{4.2}), which may be the main cause of smaller $M_*/L$ values for the DustPedia sample.

The scatter in the $M_*/L$ ratios calculated from the stellar luminosity is significantly reduced (Fig. 5). At wavelengths longer than 1$\mu$m, the overall scatter ranges between 0.1 and 0.11 dex, demonstrating a weak dependence on the wavelength. This indicates that the NIR spectrum is valuable for constraining the stellar mass almost regardless of the wavelength if the dust component can be adequately removed.  
%Similar to the $M_*/L$ ratios calculated using the total luminosity, the scatter in the $M_*/L$ calculated using the stellar luminosity from the DustPedia sample is significantly larger than those of the S82 sample. 
The $M_*/L$ values and their scatter for both subsamples are summarized in Table 1.

\section{Discussion}
\subsection{Dust Contribution to NIR Photometry}
The NIR colors (IRAC1-IRAC2 or W1-W2) have been utilized to approximate the light contribution from the dust, which needs to be adequately subtracted to estimate the stellar mass precisely. The basic assumption of this procedure is that the dust continuum is intrinsically redder than the stellar continuum. According to this assumption, \cite{querejeta2015} and \cite{jarrett2023} demonstrated that $M_*/L$ is inversely proportional to both the IRAC1-IRAC2 and W1-W2 colors, respectively. Here, we examine the validity of this finding for our dataset. We find two sequences in the plane of $M_*/L$ and W1-W2 color for our sample (Fig. 6). In one sequence, consistent with the findings of previous studies, $M_*/L$ is mildly anti-correlated with the W1-W2 and IRAC1-IRAC2 color. Although this correlation broadly agrees with the general trend in earlier studies \cite[e.g.,][]{cluver2014,jarrett2023}, the observed relation between the two quantities deviates from those of previous studies. We attribute this to the fact that the photometric measurements of W1 and W2 adopted in this study are systematically brighter than the ALLWISE measurements (\citealt{li2023}).

%previous studies (\citealp{cluver2014,jarrett2023}) utilized the ALLWISE fluxes while .} 

\begin{figure}[tbp!]
\centering
\includegraphics[width=0.45\textwidth]{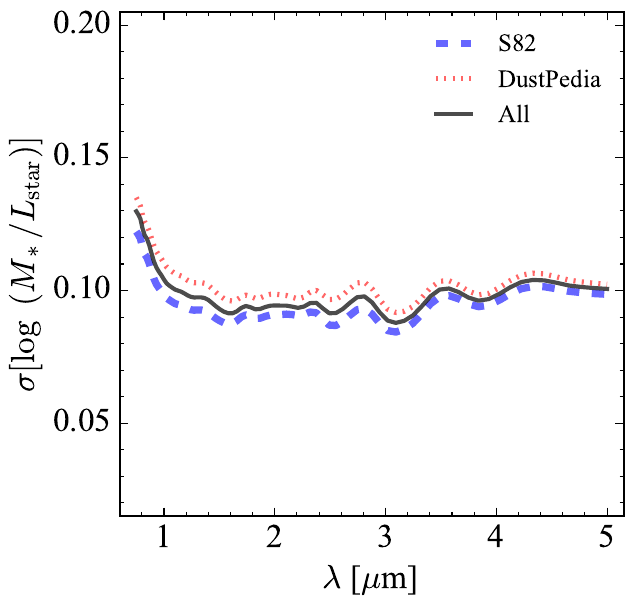}
\caption{
Scatter in $M_*/L$ after correcting the correlation between $M_*/L$ and the stellar mass ($M_*$). The symbols are the same as in Figure 4.}
\end{figure}

However, in another sequence, $M_*/L$ is almost independent of the W1$-$W2 color, with a substantially large scatter. To comprehend this finding, we compare the NIR color with the ratio of the flux from the dust components to the total flux ($f_{\rm dust}$) at the W1 band. Figure 7 shows that the dust fraction at the W1 band is only weakly correlated with the W1$-$W2 color. The hot and warm dust ($T\sim300-500$ K) from the star-forming regions can naturally increase the W1$-$W2 color \cite[e.g.,][]{xie2018}. Conversely, the prominent 3.3 $\mu$m PAH emission can boost the flux only at the W1 band, making the W1$-$W2 color bluer. Because the strength of the 3.3 $\mu$m PAH emission is proportional to the SFR but with a significant amount of the scatter, the correlation between the W1$-$W2 color and the dust fraction can be excessively intricate due to the competition between the dust continuum and PAH emission. This complexity cannot be quantified without a spectroscopic dataset \citep[e.g.,][]{yamada2013, shim2023}. A similar trend (i.e., a non-linear correlation between the dust fraction and the NIR color) is also shown in the IRAC1$-$IRAC2 color.  

As SPHEREx will provide the spectral information, the spectral color can serve as a better tracer for the dust contribution. Here, we adopt a color index between 3 $\mu$m and 4 $\mu$m ($m_{\rm 3\mu m} - m_{\rm 4\mu m}$) to avoid the strong 3.3 $\mu$m emission and compare it with the flux ratio of the dust at 3 $\mu$m ($f_{\rm dust, 3\mu m}$). We find that $f_{\rm dust, 3\mu m}$ is strongly correlated with $m_{\rm 3\mu m} - m_{\rm 4\mu m}$ expressed as $f_{\rm dust, 3\mu m}=-0.05+0.41(m_{\rm 3\mu m} - m_{\rm 4\mu m})$ (Fig. 8), indicating that the dust contribution can be approximately estimated with the spectral color index in SPHEREx dataset.

\begin{figure}[tbp!]
\centering
\includegraphics[width=0.45\textwidth]{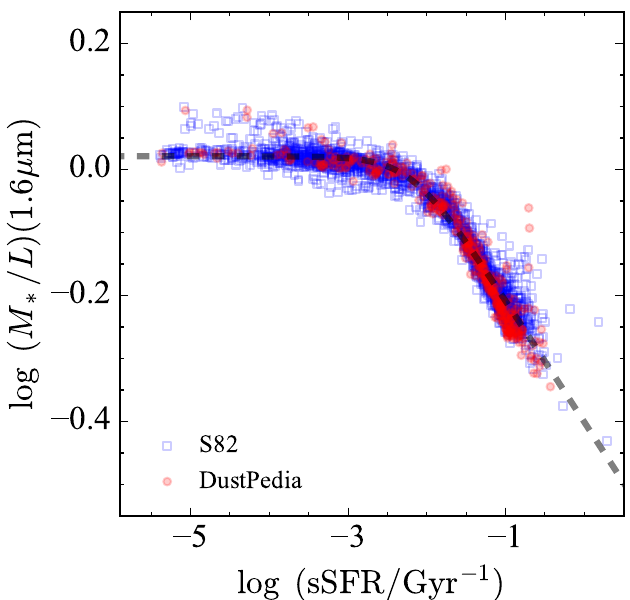}
\caption{
Correlation between $M_*/L$ and the sSFR (i.e., SFR/$M_*$) at $\sim1.6$ $\mu$m. Note that $M_*/L$ is calculated using the stellar luminosity. The symbols are the same as in Figure 1. The relation can be well fitted with the smoothly broken power-law model (dashed line).}
\end{figure}

\subsection{Main Driver of the Scatter in $M_*/L$}
Given that the $M_*/L$ ratios can be highly affected by the stellar properties of the galaxies, adopting a second parameter is often essential for robustly estimating the stellar mass. Previous studies have extensively adopted the color term as the second parameter \citep[e.g.,][]{bell2003,into2013}. To identify an alternative in the spectral analysis, we investigate the parameter predominantly deriving the scatter in $M_*/L$. Initially, we find that $M_*/L$ is weakly correlated with the stellar mass itself, likely attributed to the dependence of the SFH on stellar mass. To account for this dependence, we perform a linear regression between the stellar mass and $M_*/L$ at each wavelength as follows: $\log\ (M_*/L) = a + b\log\ (M_*/M_\odot)$ (see Figure 9 for an example). The fitting results as a function of the wavelength are summarized in Table 2. With this treatment, the scatter is mildly reduced by $\sim0.01-0.02$ dex. Thus, overall, with the NIR spectrum, the stellar mass can be estimated with a typical uncertainty of $\sim0.09-0.1$ dex (Fig. 10).       

Furthermore, the $M_*/L$ ratio is tightly correlated with the SFR divided by the stellar mass, namely the specific star formation rate (sSFR). Figure 11 shows that this relation at $\sim1.6\ \mu$m is well fitted with a smoothly broken power law, which is defined as: 
\begin{equation}
\log\ (M_*/L)= a \left\{\frac{1}{2}\left[1+\left( \frac{\log\ {\rm sSFR}}{\delta}\right)^{1/\beta}\right] \right\}^{-\alpha\beta},
\end{equation}
where $a$ is the normalization factor [i.e., $\log\ (M_*/L)$ at ${\rm sSFR} \ll \delta$], $\delta$ is the pivot sSFR, $\alpha$ is the power-law index at ${\rm sSFR} > \delta$, and $\beta$ is the smoothness parameter, which determines the smoothness of the power-law slope change around the pivot sSFR. Note that the unit of sSFR is Gyr$^{-1}$. Fitting results for various wavelengths are summarized in Table 3. This trend is rather expected from the adopted SFH with two components (old and young stellar populations) because the sSFR approximately traces the mass ratio of the old stellar population to the young stellar population. Considering this dependency, the scatter in $M_*/L$ dramatically decreases to $\sim 0.02$ dex (Fig. 12). This result clearly demonstrates that more precise stellar mass estimations can be achieved using the known SFR. Note that the SFR used in this analysis is an instantaneous SFR provided by the CIGALE based on the star formation history. As the commonly used SFR indicators (e.g., hydrogen recombination lines) trace the average SFR over the last $\sim10$ Myr, the observed SFRs may differ from those provided by CIGALE to some degree, which can introduce additional uncertainty in the $M_*/L$ estimation \cite[e.g.,][]{byun2021}. 

\begin{figure}[tbp!]
\centering
\includegraphics[width=0.45\textwidth]{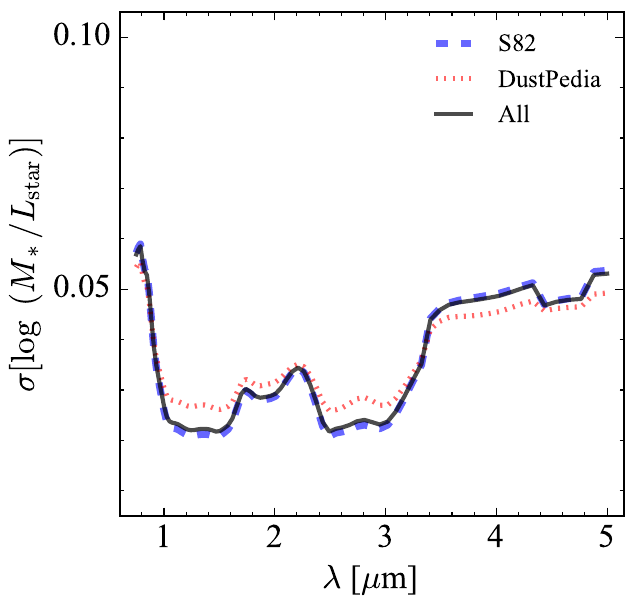}
\caption{
Scatter in $M_*/L$ after correcting the correlation between $M_*/L$ and the sSFR. The symbols are the same as in Figure 4.}
\end{figure}

\subsection{Best Stellar Mass Indicator in NIR}
Based on our calculations of $M_*/L$ under various conditions, we attempt to identify an optimal method to estimate the stellar mass based on the NIR spectral data of nearby galaxies. If the contribution of the dust emission can be adequately removed from the observed spectrum, the scatter in the $M_*/L$ ratio becomes almost independent of the wavelength. However, it is challenging to fit the PAH emissions and the dust continuum robustly, even using the NIR/MIR spectroscopic data \cite[e.g.][]{xie2018, zhang2023}. Therefore, if available, NIR data around a wavelength of $1.6\ \mu$m are preferable, as will also be presented in Lee, J. H. et al. (submitted). This is because it is relatively free from dust contamination unless the AGN component is prominent, and the resulting scatter in $M_*/L$ is smaller or comparable to that achieved using NIR data.

As outlined in the previous section (\S{4.2}), the scatter in $M_*/L$ can be slightly reduced if the dependence of $M_*/L$ on stellar mass is considered. To this end, the stellar mass can be estimated from $M_*/L$ using the monochromatic luminosity and subsequently subjected to iterative fine-tuning. Specifically, the initial stellar mass is estimated from a constant $M_*/L$, following which the correlation between the $M_*/L$ ratio and stellar mass can be corrected using the initial stellar mass measurement to achieve a more precise stellar-mass estimation. However, this iterative process can reduce the uncertainty by only $0.01-0.02$ dex. Conversely, if the SFR can be estimated based on the NIR spectral data, for example, from the fluxes of hydrogen recombination lines \cite[e.g.,][]{kennicutt2012} or 3.3 $\mu$m PAH emissions \cite[e.g.][]{kim2012, lai2020, belfiore2023b}, the scatter in the stellar mass estimation can be dramatically reduced based on the strong correlation between $M_*/L$ and sSFR. It is worthwhile to note that 3.3 $\mu$m PAH-based SFR can be more uncertain compared to other SFR indicators (e.g., hydrogen lines, UV, and IR) as its brightness relative to IR luminosity is correlated with physical parameters of host galaxies, such as metallicity \cite[e.g.][]{shim2023, whitcomb2024}. However, this correction must again follow an iterative procedure, as a prior stellar mass estimate is essential to compute the sSFR.  
An alternative argument is that SFR, rather than sSFR, could be used to trace $M_*/L$. However, we find that the scatter in the correlation between SFR and $M_*/L$ is significantly larger than that for sSFR. Therefore, sSFR provides a more reliable estimate of $M_*/L$. In conclusion, the NIR spectrophotometry provided by the SPHEREx can improve the accuracy of stellar mass estimations.

\section{Conclusion}
To investigate the accuracy of the $M_*/L$ ratio at the NIR continuum, we employ the synthetic SEDs of stellar populations and dust derived from the observed multi-wavelength data ranging from UV to FIR of nearby galaxies at $z<0.11$. The SED fitting results for two subsamples from the DustPedia and the S82 regions are adopted for this purpose. The stellar masses of the sample are also calculated based on the SED fitting results. The followings are some key findings and conclusions of our study:  
\begin{itemize}
\item The $M_*/L$ ratios in the widely used NIR broadband filters (e.g., W1 and IRAC1) are heavily affected by the strength of the dust continuum and PAH emissions. Our results demonstrate that the variations in the $M_*/L$ ratio resulting from this effect cannot be easily quantified using the NIR color (e.g., W1$-$W2 or IRAC1$-$IRAC2) because the relative contribution between the dust continuum and PAH emission is complicated. This systematics effect can be significantly alleviated using the color index of NIR spectral data. 

\item While $M_*/L$ in the NIR spectral range is strongly correlated with the wavelength, the scatter in $M_*/L$ as a function of the wavelength exhibits a U-shaped profile due to the dependence on the stellar age and dust features in the short and long wavelength, respectively. 
\item If the contribution from the dust component can be adequately excluded from the NIR spectrum, the scatter in $M_*/L$ can be significantly reduced in the long wavelength region and remain nearly constant along the wavelength with a variation of $0.01-0.02$ dex. 
%Our results reveal that the scatter in $M_*/L$ is sensitive to the adopted SFH. In particular, the flexible SFH employed to fit the DustPedia sample results in a significantly larger scatter (by $\sim0.02-0.03$ dex) compared to the double exponential SFH adopted for the massive galaxies from the S82.
\item $M_*/L$ demonstrates a weak correlation with the stellar mass and a strong correlation with the sSFR. If this dependency is corrected appropriately, the scatter in $M_*/L$ can be dramatically decreased up to $0.02-0.06$ dex. Based on these findings, we conclude that the stellar mass can be most accurately estimated using spectral data around $\sim 1.6 \mu$m, combined with SFR measurements. 
\end{itemize}
%%% ACKNOWLEDGMENTS (IF ANY) %%%%%%%%%%%%%%%%%%%%%%%%%%%%%%%%%%%%%%%%

\begin{acknowledgments}
We are grateful to the anonymous referee for constructive comments and suggestions that greatly improved our manuscript. This work was supported by the National Key R\&D Program of China (2022YFF0503401), the National Science Foundation of China (11991052, 12233001), the China Manned Space Project (CMS-CSST-2021-A04, CMS-CSST-2021-A06), and the National Research Foundation of Korea (NRF), through grants funded by the Korean government (MSIT) (Nos. 2022R1A4A3031306, 2023R1A2C1006261, RS-2023-00240212, and RS-2024-00347548). Y. K. was supported by the National Research Foundation of Korea (NRF) grant funded by the Korean government (MSIT) (No. 2021R1C1C2091550). B. L. was supported by the National Research Foundation of Korea(NRF) grant funded by the Korean government(MSIT), No. 2022R1C1C1008695. D.K. acknowledges the support of the National Research Foundation of Korea (NRF) grant (No. 2021R1C1C1013580). J.H.L was supported by Basic Science Research Program through the National Research Foundation of Korea (NRF) funded by the Ministry of Education (No. RS-2024-00452816).

\end{acknowledgments}

\appendix
\section{Comparison with Previous Studies for the DustPedia Sample}
For the DustPedia sample, the SED fitting with distinct fitting parameters was performed in the previous study (\citealp{nersesian2019}). Specifically, the DustPedia sample was fitted with a flexible-delayed SFH, comprising both young ($t_{\rm age} \leq 200$ Myr) and old ($t_{\rm age} > 200$ Myr) SSPs in \cite{nersesian2019}. Among these SSPs, the old stellar population with an age range between 2 and 12 Gyr was assumed to exhibit an exponentially declining SFR, while the young stellar population was modeled with a burst or a decline of constant SFR at 200 Myr ago. The IMF of \citet[][]{salpeter1955} was adopted. In addition, the dust emission was modeled using a broader range of parameters than in this study. In this appendix, we investigate systematics arising from differences in SFH and dust modeling by comparing our fitting results with those from \cite{nersesian2019}.
The discrepancy in the stellar masses is nearly negligible on average [$\Delta \log\ (M_*/M_\odot) \equiv \log\ (M_*/M_\odot) ({\rm this\ study}) - \log\ (M_*/M_\odot) ({\rm Nersesian}) \sim 0.01$] with a scatter of $\sim0.12$ dex, which may correspond to the additional uncertainty introduced by differences in SFH and IMF in the SED fitting (Fig. A1). The SFR shows a marginally larger offset and scatter [$\log\ ({\rm SFR}/M_\odot {\rm yr^{-1}}) = -0.10 \pm 0.33$ dex; Fig. A1]. However, this uncertainty will have minimal impact on this study.

\begin{figure*}[t!]
\centering
\figurenum{A1}
\includegraphics[width=0.45\textwidth]{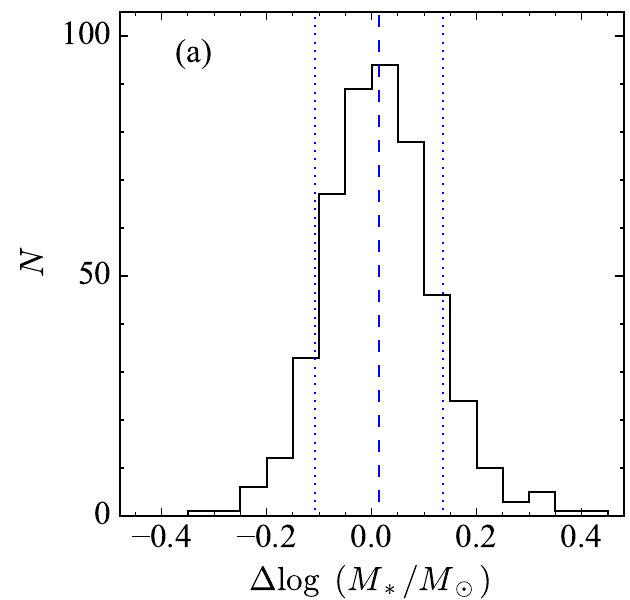}
\includegraphics[width=0.45\textwidth]{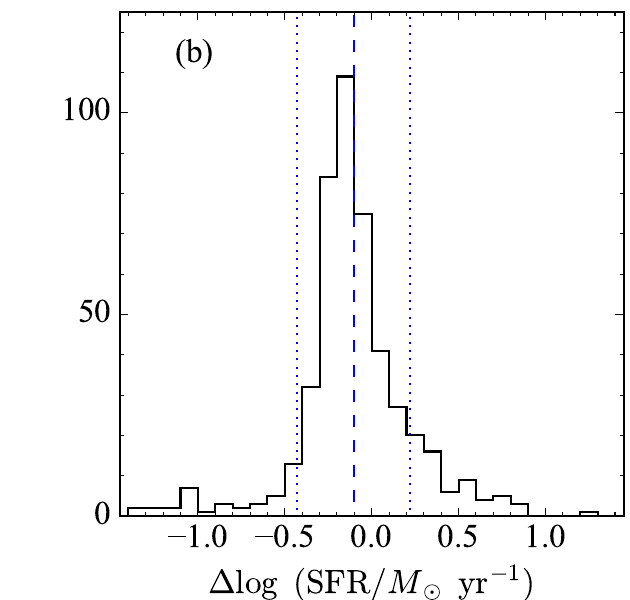}
\caption{
Histograms of differences in the stellar masses (a) and SFR (b) of the DustPedia sample between this study and \cite{nersesian2019}. The dashed and dotted lines denote the mean value and standard deviation, respectively.}
\end{figure*}

\section{Residuals from the SED Fitting in NIR bands}
The SED model in the NIR is highly sensitive to the choice of stellar population model, particularly based on how TP-AGB stars are treated \cite[e.g.,][]{maraston2006}. To understand the systematics of $M_*/L$ due to this fact, it is helpful to assess the goodness of fit by examining the NIR residuals. Figure A2 illustrates the residuals in the NIR bands ($JHK_s$, W1, and W2) derived from the best SED fit. Note only the results with a signal-to-noise greater than 5 are used in the experiment. Notably, the model SEDs tend to underestimate the fluxes at $JHK_s$, though the offsets remain within the error margin. This indicates that the NIR-based $M_*/L$ could be marginally underestimated up to $\sim0.04$ dex, although this result is not definitive.

\begin{figure}[tbp!]
\centering
\figurenum{A2}
\includegraphics[width=0.45\textwidth]{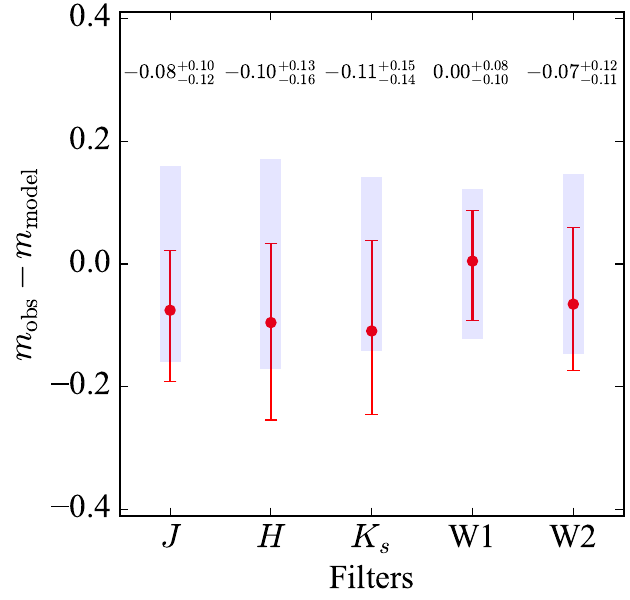}
\caption{
Distributions of residual magnitudes between the observed magnitudes ($m_{\rm obs}$) and the model magnitudes ($m_{\rm model}$) derived from the best-fit SED in the NIR bands ($JHK_s$, W1, and W2). For each filter, the median and 16/84th percentiles are displayed with the shaded region representing the typical uncertainties in the observed magnitudes.}
\end{figure}

\bibliography{m_l}

\begin{thebibliography}{}
\expandafter\ifx\csname natexlab\endcsname\relax\def\natexlab#1{#1}\fi

\bibitem[{{Belfiore} {et~al.}(2023){Belfiore}, {Leroy}, {Williams}, {Barnes}, {Bigiel}, {Boquien}, {Cao}, {Chastenet}, {Congiu}, {Dale}, {Egorov}, {Eibensteiner}, {Emsellem}, {Glover}, {Groves}, {Hassani}, {Klessen}, {Kreckel}, {Neumann}, {Neumann}, {Querejeta}, {Rosolowsky}, {Sanchez-Blazquez}, {Sandstrom}, {Schinnerer}, {Sun}, {Sutter}, \& {Watkins}}]{belfiore2023b}
{Belfiore}, F., {Leroy}, A.~K., {Williams}, T.~G., {et~al.} 2023, \aap, 678, A129

\bibitem[{{Bell} {et~al.}(2003){Bell}, {McIntosh}, {Katz}, \& {Weinberg}}]{bell2003}
{Bell}, E.~F., {McIntosh}, D.~H., {Katz}, N., \& {Weinberg}, M.~D. 2003, \apjs, 149, 289

\bibitem[{{Boquien} {et~al.}(2019){Boquien}, {Burgarella}, {Roehlly}, {Buat}, {Ciesla}, {Corre}, {Inoue}, \& {Salas}}]{boquien2019}
{Boquien}, M., {Burgarella}, D., {Roehlly}, Y., {et~al.} 2019, \aap, 622, A103

\bibitem[{{Brown} {et~al.}(2014){Brown}, {Moustakas}, {Smith}, {da Cunha}, {Jarrett}, {Imanishi}, {Armus}, {Brandl}, \& {Peek}}]{brown2014}
{Brown}, M. J.~I., {Moustakas}, J., {Smith}, J. D.~T., {et~al.} 2014, \apjs, 212, 18

\bibitem[{{Bruzual} \& {Charlot}(2003)}]{bruzual2003}
{Bruzual}, G., \& {Charlot}, S. 2003, \mnras, 344, 1000

\bibitem[{{Byun} {et~al.}(2021){Byun}, {Sheen}, {Seon}, {Ho}, {Lee}, {Jeong}, {Kim}, {Park}, {Lee}, {Cha}, {Ko}, \& {Kim}}]{byun2021}
{Byun}, W., {Sheen}, Y.-K., {Seon}, K.-I., {et~al.} 2021, \apj, 918, 82

\bibitem[{{Calzetti} {et~al.}(2000){Calzetti}, {Armus}, {Bohlin}, {Kinney}, {Koornneef}, \& {Storchi-Bergmann}}]{calzetti2000}
{Calzetti}, D., {Armus}, L., {Bohlin}, R.~C., {et~al.} 2000, \apj, 533, 682

\bibitem[{{Chabrier}(2003)}]{chabrier2003}
{Chabrier}, G. 2003, \pasp, 115, 763

\bibitem[{{Clark} {et~al.}(2018){Clark}, {Verstocken}, {Bianchi}, {Fritz}, {Viaene}, {Smith}, {Baes}, {Casasola}, {Cassara}, {Davies}, {De Looze}, {De Vis}, {Evans}, {Galametz}, {Jones}, {Lianou}, {Madden}, {Mosenkov}, \& {Xilouris}}]{clark2018}
{Clark}, C.~J.~R., {Verstocken}, S., {Bianchi}, S., {et~al.} 2018, \aap, 609, A37

\bibitem[{{Cluver} {et~al.}(2014){Cluver}, {Jarrett}, {Hopkins}, {Driver}, {Liske}, {Gunawardhana}, {Taylor}, {Robotham}, {Alpaslan}, {Baldry}, {Brown}, {Peacock}, {Popescu}, {Tuffs}, {Bauer}, {Bland-Hawthorn}, {Colless}, {Holwerda}, {Lara-L{\'o}pez}, {Leschinski}, {L{\'o}pez-S{\'a}nchez}, {Norberg}, {Owers}, {Wang}, \& {Wilkins}}]{cluver2014}
{Cluver}, M.~E., {Jarrett}, T.~H., {Hopkins}, A.~M., {et~al.} 2014, \apj, 782, 90

\bibitem[{{Conroy}(2013)}]{conroy2013}
{Conroy}, C. 2013, \araa, 51, 393

\bibitem[{{Conselice}(2014)}]{conselice2014}
{Conselice}, C.~J. 2014, \araa, 52, 291

\bibitem[{{Crill} {et~al.}(2020){Crill}, {Werner}, {Akeson}, {Ashby}, {Bleem}, {Bock}, {Bryan}, {Burnham}, {Byunh}, {Chang}, {Chiang}, {Cook}, {Cooray}, {Davis}, {Dor{\'e}}, {Dowell}, {Dubois-Felsmann}, {Eifler}, {Faisst}, {Habib}, {Heinrich}, {Heitmann}, {Heaton}, {Hirata}, {Hristov}, {Hui}, {Jeong}, {Kang}, {Kecman}, {Kirkpatrick}, {Korngut}, {Krause}, {Lee}, {Lisse}, {Masters}, {Mauskopf}, {Melnick}, {Miyasaka}, {Nayyeri}, {Nguyen}, {{\"O}berg}, {Padin}, {Paladini}, {Pourrahmani}, {Pyo}, {Smith}, {Song}, {Symons}, {Teplitz}, {Tolls}, {Unwin}, {Windhorst}, {Yang}, \& {Zemcov}}]{crill2020}
{Crill}, B.~P., {Werner}, M., {Akeson}, R., {et~al.} 2020, in Society of Photo-Optical Instrumentation Engineers (SPIE) Conference Series, Vol. 11443, Space Telescopes and Instrumentation 2020: Optical, Infrared, and Millimeter Wave, ed. M.~{Lystrup} \& M.~D. {Perrin}, 114430I

\bibitem[{{Cutri} {et~al.}(2003){Cutri}, {Skrutskie}, {van Dyk}, {Beichman}, {Carpenter}, {Chester}, {Cambresy}, {Evans}, {Fowler}, {Gizis}, {Howard}, {Huchra}, {Jarrett}, {Kopan}, {Kirkpatrick}, {Light}, {Marsh}, {McCallon}, {Schneider}, {Stiening}, {Sykes}, {Weinberg}, {Wheaton}, {Wheelock}, \& {Zacarias}}]{cutri2003}
{Cutri}, R.~M., {Skrutskie}, M.~F., {van Dyk}, S., {et~al.} 2003, {2MASS All Sky Catalog of point sources.}

\bibitem[{{Davies} {et~al.}(2017){Davies}, {Baes}, {Bianchi}, {Jones}, {Madden}, {Xilouris}, {Bocchio}, {Casasola}, {Cassara}, {Clark}, {De Looze}, {Evans}, {Fritz}, {Galametz}, {Galliano}, {Lianou}, {Mosenkov}, {Smith}, {Verstocken}, {Viaene}, {Vika}, {Wagle}, \& {Ysard}}]{davies2017}
{Davies}, J.~I., {Baes}, M., {Bianchi}, S., {et~al.} 2017, \pasp, 129, 044102

\bibitem[{{Dor{\'e}} {et~al.}(2016){Dor{\'e}}, {Werner}, {Ashby}, {Banerjee}, {Battaglia}, {Bauer}, {Benjamin}, {Bleem}, {Bock}, {Boogert}, {Bull}, {Capak}, {Chang}, {Chiar}, {Cohen}, {Cooray}, {Crill}, {Cushing}, {de Putter}, {Driver}, {Eifler}, {Feng}, {Ferraro}, {Finkbeiner}, {Gaudi}, {Greene}, {Hillenbrand}, {H{\"o}flich}, {Hsiao}, {Huffenberger}, {Jansen}, {Jeong}, {Joshi}, {Kim}, {Kim}, {Kirkpatrick}, {Korngut}, {Krause}, {Kriek}, {Leistedt}, {Li}, {Lisse}, {Mauskopf}, {Mechtley}, {Melnick}, {Mohr}, {Murphy}, {Neben}, {Neufeld}, {Nguyen}, {Pierpaoli}, {Pyo}, {Rhodes}, {Sandstrom}, {Schaan}, {Schlaufman}, {Silverman}, {Su}, {Stassun}, {Stevens}, {Strauss}, {Tielens}, {Tsai}, {Tolls}, {Unwin}, {Viero}, {Windhorst}, \& {Zemcov}}]{dore2016}
{Dor{\'e}}, O., {Werner}, M.~W., {Ashby}, M., {et~al.} 2016, arXiv e-prints, arXiv:1606.07039

\bibitem[{{Fritz} {et~al.}(2006){Fritz}, {Franceschini}, \& {Hatziminaoglou}}]{fritz2006}
{Fritz}, J., {Franceschini}, A., \& {Hatziminaoglou}, E. 2006, \mnras, 366, 767

\bibitem[{{Hon} {et~al.}(2022){Hon}, {Graham}, {Davis}, \& {Marconi}}]{hon2022}
{Hon}, D. S.~H., {Graham}, A.~W., {Davis}, B.~L., \& {Marconi}, A. 2022, \mnras, 514, 3410

\bibitem[{{Inami} {et~al.}(2018){Inami}, {Armus}, {Matsuhara}, {Charmandaris}, {D{\'\i}az-Santos}, {Surace}, {Stierwalt}, {Ohyama}, {Howell}, {Marshall}, {Evans}, {Linden}, \& {Mazzarella}}]{inami2018}
{Inami}, H., {Armus}, L., {Matsuhara}, H., {et~al.} 2018, \aap, 617, A130

\bibitem[{{Inoue}(2011)}]{inoue2011}
{Inoue}, A.~K. 2011, \mnras, 415, 2920

\bibitem[{{Into} \& {Portinari}(2013)}]{into2013}
{Into}, T., \& {Portinari}, L. 2013, \mnras, 430, 2715

\bibitem[{{Jarrett} {et~al.}(2000){Jarrett}, {Chester}, {Cutri}, {Schneider}, {Skrutskie}, \& {Huchra}}]{jarrett2000}
{Jarrett}, T.~H., {Chester}, T., {Cutri}, R., {et~al.} 2000, \aj, 119, 2498

\bibitem[{{Jarrett} {et~al.}(2023){Jarrett}, {Cluver}, {Taylor}, {Bellstedt}, {Robotham}, \& {Yao}}]{jarrett2023}
{Jarrett}, T.~H., {Cluver}, M.~E., {Taylor}, E.~N., {et~al.} 2023, \apj, 946, 95

\bibitem[{{Jones} {et~al.}(2013){Jones}, {Fanciullo}, {K{\"o}hler}, {Verstraete}, {Guillet}, {Bocchio}, \& {Ysard}}]{jones2013}
{Jones}, A.~P., {Fanciullo}, L., {K{\"o}hler}, M., {et~al.} 2013, \aap, 558, A62

\bibitem[{{Kennicutt} \& {Evans}(2012)}]{kennicutt2012}
{Kennicutt}, R.~C., \& {Evans}, N.~J. 2012, \araa, 50, 531

\bibitem[{{Kettlety} {et~al.}(2018){Kettlety}, {Hesling}, {Phillipps}, {Bremer}, {Cluver}, {Taylor}, {Bland-Hawthorn}, {Brough}, {De Propris}, {Driver}, {Holwerda}, {Kelvin}, {Sutherland}, \& {Wright}}]{kettlety2018}
{Kettlety}, T., {Hesling}, J., {Phillipps}, S., {et~al.} 2018, \mnras, 473, 776

\bibitem[{{Kim} {et~al.}(2012){Kim}, {Im}, {Lee}, {Lee}, {Jun}, {Nakagawa}, {Matsuhara}, {Wada}, {Oyabu}, {Takagi}, {Inami}, {Ohyama}, {Yamada}, {Helou}, {Armus}, \& {Shi}}]{kim2012}
{Kim}, J.~H., {Im}, M., {Lee}, H.~M., {et~al.} 2012, \apj, 760, 120

\bibitem[{{Kormendy} \& {Ho}(2013)}]{kormendy2013}
{Kormendy}, J., \& {Ho}, L.~C. 2013, \araa, 51, 511

\bibitem[{{Korngut} {et~al.}(2018){Korngut}, {Bock}, {Akeson}, {Ashby}, {Bleem}, {Boland}, {Bolton}, {Bradford}, {Braun}, {Bryan}, {Capak}, {Chang}, {Coffey}, {Cooray}, {Crill}, {Dor{\'e}}, {Eifler}, {Feng}, {Habib}, {Heitmann}, {Hemmati}, {Hirata}, {Jeong}, {Kim}, {Kirkpatrick}, {Kowalkowski}, {Krause}, {Lisse}, {Mauskopf}, {Masters}, {McGuire}, {Melnick}, {Nguyen}, {Nayyeri}, {Oberg}, {dePutter}, {Purcell}, {Rocca}, {Runyan}, {Sandstrom}, {Smith}, {Song}, {Stickley}, {Stober}, {Susca}, {Teplitz}, {Tolls}, {Unwin}, {Werner}, {Windhorst}, \& {Zemcov}}]{korngut2018}
{Korngut}, P.~M., {Bock}, J.~J., {Akeson}, R., {et~al.} 2018, in Society of Photo-Optical Instrumentation Engineers (SPIE) Conference Series, Vol. 10698, Space Telescopes and Instrumentation 2018: Optical, Infrared, and Millimeter Wave, ed. M.~{Lystrup}, H.~A. {MacEwen}, G.~G. {Fazio}, N.~{Batalha}, N.~{Siegler}, \& E.~C. {Tong}, 106981U

\bibitem[{{Lai} {et~al.}(2020){Lai}, {Smith}, {Baba}, {Spoon}, \& {Imanishi}}]{lai2020}
{Lai}, T. S.~Y., {Smith}, J.~D.~T., {Baba}, S., {Spoon}, H. W.~W., \& {Imanishi}, M. 2020, \apj, 905, 55

\bibitem[{{Lang} {et~al.}(2016){Lang}, {Hogg}, \& {Schlegel}}]{lang2016}
{Lang}, D., {Hogg}, D.~W., \& {Schlegel}, D.~J. 2016, \aj, 151, 36

\bibitem[{{Larson} \& {Tinsley}(1978)}]{larson1978}
{Larson}, R.~B., \& {Tinsley}, B.~M. 1978, \apj, 219, 46

\bibitem[{{Lee} {et~al.}(2012){Lee}, {Hwang}, {Lee}, {Kim}, \& {Lee}}]{lee2012}
{Lee}, J.~C., {Hwang}, H.~S., {Lee}, M.~G., {Kim}, M., \& {Lee}, J.~H. 2012, \apj, 756, 95

\bibitem[{{Li} {et~al.}(2023){Li}, {Ho}, {Shangguan}, {Zhuang}, \& {Li}}]{li2023}
{Li}, Y.~A., {Ho}, L.~C., {Shangguan}, J., {Zhuang}, M.-Y., \& {Li}, R. 2023, \apjs, 267, 17

\bibitem[{{Lower} {et~al.}(2020){Lower}, {Narayanan}, {Leja}, {Johnson}, {Conroy}, \& {Dav{\'e}}}]{lower2020}
{Lower}, S., {Narayanan}, D., {Leja}, J., {et~al.} 2020, \apj, 904, 33

\bibitem[{{Madau} \& {Dickinson}(2014)}]{madau2014}
{Madau}, P., \& {Dickinson}, M. 2014, \araa, 52, 415

\bibitem[{{Maiolino} \& {Mannucci}(2019)}]{maiolino2019}
{Maiolino}, R., \& {Mannucci}, F. 2019, \aapr, 27, 3

\bibitem[{{Maraston} {et~al.}(2006){Maraston}, {Daddi}, {Renzini}, {Cimatti}, {Dickinson}, {Papovich}, {Pasquali}, \& {Pirzkal}}]{maraston2006}
{Maraston}, C., {Daddi}, E., {Renzini}, A., {et~al.} 2006, \apj, 652, 85

\bibitem[{{Meidt} {et~al.}(2012){Meidt}, {Schinnerer}, {Knapen}, {Bosma}, {Athanassoula}, {Sheth}, {Buta}, {Zaritsky}, {Laurikainen}, {Elmegreen}, {Elmegreen}, {Gadotti}, {Salo}, {Regan}, {Ho}, {Madore}, {Hinz}, {Skibba}, {Gil de Paz}, {Mu{\~n}oz-Mateos}, {Men{\'e}ndez-Delmestre}, {Seibert}, {Kim}, {Mizusawa}, {Laine}, \& {Comer{\'o}n}}]{meidt2012}
{Meidt}, S.~E., {Schinnerer}, E., {Knapen}, J.~H., {et~al.} 2012, \apj, 744, 17

\bibitem[{{Meidt} {et~al.}(2014){Meidt}, {Schinnerer}, {van de Ven}, {Zaritsky}, {Peletier}, {Knapen}, {Sheth}, {Regan}, {Querejeta}, {Mu{\~n}oz-Mateos}, {Kim}, {Hinz}, {Gil de Paz}, {Athanassoula}, {Bosma}, {Buta}, {Cisternas}, {Ho}, {Holwerda}, {Skibba}, {Laurikainen}, {Salo}, {Gadotti}, {Laine}, {Erroz-Ferrer}, {Comer{\'o}n}, {Men{\'e}ndez-Delmestre}, {Seibert}, \& {Mizusawa}}]{meidt2014}
{Meidt}, S.~E., {Schinnerer}, E., {van de Ven}, G., {et~al.} 2014, \apj, 788, 144

\bibitem[{{Nersesian} {et~al.}(2019){Nersesian}, {Xilouris}, {Bianchi}, {Galliano}, {Jones}, {Baes}, {Casasola}, {Cassar{\`a}}, {Clark}, {Davies}, {Decleir}, {Dobbels}, {De Looze}, {De Vis}, {Fritz}, {Galametz}, {Madden}, {Mosenkov}, {Tr{\v{c}}ka}, {Verstocken}, {Viaene}, \& {Lianou}}]{nersesian2019}
{Nersesian}, A., {Xilouris}, E.~M., {Bianchi}, S., {et~al.} 2019, \aap, 624, A80

\bibitem[{{Norris} {et~al.}(2016){Norris}, {Van de Ven}, {Schinnerer}, {Crain}, {Meidt}, {Groves}, {Bower}, {Furlong}, {Schaller}, {Schaye}, \& {Theuns}}]{norris2016}
{Norris}, M.~A., {Van de Ven}, G., {Schinnerer}, E., {et~al.} 2016, \apj, 832, 198

\bibitem[{{Portinari} {et~al.}(2004){Portinari}, {Sommer-Larsen}, \& {Tantalo}}]{portinari2004}
{Portinari}, L., {Sommer-Larsen}, J., \& {Tantalo}, R. 2004, \mnras, 347, 691

\bibitem[{{Querejeta} {et~al.}(2015){Querejeta}, {Meidt}, {Schinnerer}, {Cisternas}, {Mu{\~n}oz-Mateos}, {Sheth}, {Knapen}, {van de Ven}, {Norris}, {Peletier}, {Laurikainen}, {Salo}, {Holwerda}, {Athanassoula}, {Bosma}, {Groves}, {Ho}, {Gadotti}, {Zaritsky}, {Regan}, {Hinz}, {Gil de Paz}, {Menendez-Delmestre}, {Seibert}, {Mizusawa}, {Kim}, {Erroz-Ferrer}, {Laine}, \& {Comer{\'o}n}}]{querejeta2015}
{Querejeta}, M., {Meidt}, S.~E., {Schinnerer}, E., {et~al.} 2015, \apjs, 219, 5

\bibitem[{{Renzini} \& {Peng}(2015)}]{renzini2015}
{Renzini}, A., \& {Peng}, Y.-j. 2015, \apjl, 801, L29

\bibitem[{{Roediger} \& {Courteau}(2015)}]{roediger2015}
{Roediger}, J.~C., \& {Courteau}, S. 2015, \mnras, 452, 3209

\bibitem[{{Salim} {et~al.}(2018){Salim}, {Boquien}, \& {Lee}}]{salim2018}
{Salim}, S., {Boquien}, M., \& {Lee}, J.~C. 2018, \apj, 859, 11

\bibitem[{{Salpeter}(1955)}]{salpeter1955}
{Salpeter}, E.~E. 1955, \apj, 121, 161

\bibitem[{{Shim} {et~al.}(2023){Shim}, {Hwang}, {Jeong}, {Toba}, {Kim}, {Kim}, {Song}, {Hashimoto}, {Nakagawa}, {Nanni}, {Pearson}, \& {Takagi}}]{shim2023}
{Shim}, H., {Hwang}, H.~S., {Jeong}, W.-S., {et~al.} 2023, \aj, 165, 31

\bibitem[{{Taylor} {et~al.}(2011){Taylor}, {Hopkins}, {Baldry}, {Brown}, {Driver}, {Kelvin}, {Hill}, {Robotham}, {Bland-Hawthorn}, {Jones}, {Sharp}, {Thomas}, {Liske}, {Loveday}, {Norberg}, {Peacock}, {Bamford}, {Brough}, {Colless}, {Cameron}, {Conselice}, {Croom}, {Frenk}, {Gunawardhana}, {Kuijken}, {Nichol}, {Parkinson}, {Phillipps}, {Pimbblet}, {Popescu}, {Prescott}, {Sutherland}, {Tuffs}, {van Kampen}, \& {Wijesinghe}}]{taylor2011}
{Taylor}, E.~N., {Hopkins}, A.~M., {Baldry}, I.~K., {et~al.} 2011, \mnras, 418, 1587

\bibitem[{{Wechsler} \& {Tinker}(2018)}]{wechsler2018}
{Wechsler}, R.~H., \& {Tinker}, J.~L. 2018, \araa, 56, 435

\bibitem[{{Werner} {et~al.}(2004){Werner}, {Roellig}, {Low}, {Rieke}, {Rieke}, {Hoffmann}, {Young}, {Houck}, {Brandl}, {Fazio}, {Hora}, {Gehrz}, {Helou}, {Soifer}, {Stauffer}, {Keene}, {Eisenhardt}, {Gallagher}, {Gautier}, {Irace}, {Lawrence}, {Simmons}, {Van Cleve}, {Jura}, {Wright}, \& {Cruikshank}}]{werner2004}
{Werner}, M.~W., {Roellig}, T.~L., {Low}, F.~J., {et~al.} 2004, \apjs, 154, 1

\bibitem[{{Whitcomb} {et~al.}(2024){Whitcomb}, {Smith}, {Sandstrom}, {Starkey}, {Donnelly}, {Draine}, {Skillman}, {Dale}, {Armus}, {Hensley}, {Lai}, \& {Kennicutt}}]{whitcomb2024}
{Whitcomb}, C.~M., {Smith}, J. D.~T., {Sandstrom}, K., {et~al.} 2024, \apj, 974, 20

\bibitem[{{Wright} {et~al.}(2010){Wright}, {Eisenhardt}, {Mainzer}, {Ressler}, {Cutri}, {Jarrett}, {Kirkpatrick}, {Padgett}, {McMillan}, {Skrutskie}, {Stanford}, {Cohen}, {Walker}, {Mather}, {Leisawitz}, {Gautier}, {McLean}, {Benford}, {Lonsdale}, {Blain}, {Mendez}, {Irace}, {Duval}, {Liu}, {Royer}, {Heinrichsen}, {Howard}, {Shannon}, {Kendall}, {Walsh}, {Larsen}, {Cardon}, {Schick}, {Schwalm}, {Abid}, {Fabinsky}, {Naes}, \& {Tsai}}]{wright2010}
{Wright}, E.~L., {Eisenhardt}, P. R.~M., {Mainzer}, A.~K., {et~al.} 2010, \aj, 140, 1868

\bibitem[{{Xie} {et~al.}(2018{\natexlab{a}}){Xie}, {Ho}, {Li}, \& {Shangguan}}]{xie2018}
{Xie}, Y., {Ho}, L.~C., {Li}, A., \& {Shangguan}, J. 2018{\natexlab{a}}, \apj, 860, 154

\bibitem[{{Xie} {et~al.}(2018{\natexlab{b}}){Xie}, {Ho}, {Li}, \& {Shangguan}}]{xie2018b}
---. 2018{\natexlab{b}}, \apj, 867, 91

\bibitem[{{Yamada} {et~al.}(2013){Yamada}, {Oyabu}, {Kaneda}, {Yamagishi}, {Ishihara}, {Kim}, \& {Im}}]{yamada2013}
{Yamada}, R., {Oyabu}, S., {Kaneda}, H., {et~al.} 2013, \pasj, 65, 103

\bibitem[{{Zhang} \& {Ho}(2023)}]{zhang2023}
{Zhang}, L., \& {Ho}, L.~C. 2023, \apj, 943, 60

\bibitem[{{Zhang} {et~al.}(2021){Zhang}, {Ho}, \& {Xie}}]{zhang2021}
{Zhang}, L., {Ho}, L.~C., \& {Xie}, Y. 2021, \aj, 161, 29

\bibitem[{{Zibetti} {et~al.}(2009){Zibetti}, {Charlot}, \& {Rix}}]{zibetti2009}
{Zibetti}, S., {Charlot}, S., \& {Rix}, H.-W. 2009, \mnras, 400, 1181

\end{thebibliography}

\startlongtable
\begin{deluxetable*}{@{\extracolsep{4pt}}cccccccccccccc}
\tablecaption{Mass-to-light ratio in NIR Spectrum\label{tab:table1}}
\tablewidth{0pc}
\tabletypesize{\scriptsize}
\tablehead{ 
\colhead{} &
\multicolumn{6}{c}{Total Luminosity} &
 &
\multicolumn{6}{c}{Stellar Luminosity} \\
\cline{2-7} \cline{9-14} 
& \multicolumn{2}{c}{DustPedia} &
\multicolumn{2}{c}{S82} &
\multicolumn{2}{c}{All} & &
\multicolumn{2}{c}{DustPedia} &
\multicolumn{2}{c}{S82} &
\multicolumn{2}{c}{All} \\
\cline{2-3} 
\cline{4-5} 
\cline{6-7} 
\cline{9-10} 
\cline{11-12} 
\cline{13-14}  
\colhead{$\lambda_c$} & 
\colhead{$\log\ (M_*/L)$} & \colhead{$\sigma$} & 
\colhead{$\log\ (M_*/L)$} & \colhead{$\sigma$} & 
\colhead{$\log\ (M_*/L)$} & \colhead{$\sigma$} & \colhead{}& 
\colhead{$\log\ (M_*/L)$} & \colhead{$\sigma$} & 
\colhead{$\log\ (M_*/L)$} & \colhead{$\sigma$} & 
\colhead{$\log\ (M_*/L)$} & \colhead{$\sigma$} \\
\colhead{(1)} & \colhead{(2)} & \colhead{(3)} & \colhead{(4)} & \colhead{(5)} & 
\colhead{(6)} & \colhead{(7)} & \colhead{} & \colhead{(8)} & \colhead{(9)} & \colhead{(10)} & 
\colhead{(11)} & \colhead{(12)} & \colhead{(13)} 
}
\startdata
0.75 & $0.125$ & 0.158 & $0.190$ & 0.134 & $0.180$ & 0.140 &  & $0.074$ & 0.152 & $0.132$ & 0.147 & $0.123$ & 0.149 \\
0.77 & $0.116$ & 0.156 & $0.180$ & 0.133 & $0.170$ & 0.139 &  & $0.067$ & 0.150 & $0.125$ & 0.146 & $0.116$ & 0.148 \\
0.79 & $0.107$ & 0.154 & $0.169$ & 0.132 & $0.160$ & 0.137 &  & $0.060$ & 0.148 & $0.117$ & 0.144 & $0.108$ & 0.146 \\
0.81 & $0.094$ & 0.149 & $0.154$ & 0.128 & $0.145$ & 0.133 &  & $0.049$ & 0.144 & $0.104$ & 0.140 & $0.095$ & 0.142 \\
0.83 & $0.082$ & 0.146 & $0.140$ & 0.125 & $0.132$ & 0.130 &  & $0.038$ & 0.141 & $0.092$ & 0.137 & $0.084$ & 0.139 \\
0.85 & $0.071$ & 0.144 & $0.129$ & 0.124 & $0.121$ & 0.129 &  & $0.030$ & 0.140 & $0.083$ & 0.135 & $0.075$ & 0.137 \\
0.87 & $0.059$ & 0.141 & $0.116$ & 0.122 & $0.107$ & 0.126 &  & $0.020$ & 0.137 & $0.072$ & 0.132 & $0.064$ & 0.134 \\
0.89 & $0.046$ & 0.137 & $0.101$ & 0.118 & $0.093$ & 0.123 &  & $0.009$ & 0.134 & $0.060$ & 0.128 & $0.052$ & 0.131 \\
0.91 & $0.035$ & 0.134 & $0.089$ & 0.116 & $0.081$ & 0.120 &  & $0.000$ & 0.130 & $0.050$ & 0.125 & $0.043$ & 0.127 \\
0.93 & $0.027$ & 0.132 & $0.079$ & 0.114 & $0.071$ & 0.118 &  & $-0.006$ & 0.127 & $0.043$ & 0.122 & $0.035$ & 0.124 \\
0.95 & $0.018$ & 0.130 & $0.070$ & 0.112 & $0.062$ & 0.117 &  & $-0.012$ & 0.125 & $0.036$ & 0.120 & $0.029$ & 0.122 \\
0.98 & $0.012$ & 0.127 & $0.062$ & 0.110 & $0.054$ & 0.114 &  & $-0.018$ & 0.124 & $0.029$ & 0.118 & $0.022$ & 0.120 \\
1.00 & $0.005$ & 0.124 & $0.054$ & 0.108 & $0.047$ & 0.112 &  & $-0.024$ & 0.121 & $0.022$ & 0.116 & $0.015$ & 0.118 \\
1.02 & $-0.001$ & 0.122 & $0.046$ & 0.106 & $0.039$ & 0.110 &  & $-0.029$ & 0.119 & $0.016$ & 0.114 & $0.009$ & 0.116 \\
1.05 & $-0.008$ & 0.121 & $0.039$ & 0.106 & $0.032$ & 0.110 &  & $-0.034$ & 0.118 & $0.011$ & 0.113 & $0.004$ & 0.115 \\
1.08 & $-0.015$ & 0.121 & $0.032$ & 0.106 & $0.025$ & 0.110 &  & $-0.038$ & 0.117 & $0.007$ & 0.112 & $-0.000$ & 0.114 \\
1.10 & $-0.018$ & 0.120 & $0.029$ & 0.106 & $0.022$ & 0.109 &  & $-0.040$ & 0.116 & $0.004$ & 0.111 & $-0.002$ & 0.113 \\
1.13 & $-0.020$ & 0.118 & $0.026$ & 0.104 & $0.019$ & 0.108 &  & $-0.043$ & 0.116 & $0.001$ & 0.111 & $-0.005$ & 0.112 \\
1.16 & $-0.025$ & 0.117 & $0.020$ & 0.104 & $0.013$ & 0.107 &  & $-0.047$ & 0.115 & $-0.003$ & 0.110 & $-0.010$ & 0.112 \\
1.18 & $-0.030$ & 0.115 & $0.014$ & 0.103 & $0.008$ & 0.106 &  & $-0.051$ & 0.114 & $-0.008$ & 0.109 & $-0.014$ & 0.111 \\
1.21 & $-0.035$ & 0.114 & $0.009$ & 0.103 & $0.002$ & 0.106 &  & $-0.055$ & 0.113 & $-0.012$ & 0.108 & $-0.019$ & 0.110 \\
1.24 & $-0.042$ & 0.114 & $0.002$ & 0.103 & $-0.005$ & 0.106 &  & $-0.059$ & 0.112 & $-0.017$ & 0.107 & $-0.024$ & 0.109 \\
1.27 & $-0.050$ & 0.114 & $-0.006$ & 0.103 & $-0.013$ & 0.106 &  & $-0.065$ & 0.112 & $-0.023$ & 0.107 & $-0.030$ & 0.109 \\
1.30 & $-0.057$ & 0.114 & $-0.013$ & 0.103 & $-0.020$ & 0.106 &  & $-0.072$ & 0.112 & $-0.030$ & 0.107 & $-0.036$ & 0.109 \\
1.33 & $-0.063$ & 0.113 & $-0.020$ & 0.103 & $-0.026$ & 0.106 &  & $-0.078$ & 0.112 & $-0.036$ & 0.107 & $-0.042$ & 0.109 \\
1.37 & $-0.069$ & 0.113 & $-0.026$ & 0.103 & $-0.033$ & 0.106 &  & $-0.084$ & 0.112 & $-0.042$ & 0.107 & $-0.048$ & 0.109 \\
1.40 & $-0.076$ & 0.112 & $-0.033$ & 0.102 & $-0.040$ & 0.105 &  & $-0.090$ & 0.111 & $-0.048$ & 0.106 & $-0.054$ & 0.108 \\
1.43 & $-0.084$ & 0.110 & $-0.042$ & 0.101 & $-0.049$ & 0.104 &  & $-0.097$ & 0.109 & $-0.056$ & 0.105 & $-0.062$ & 0.107 \\
1.47 & $-0.096$ & 0.108 & $-0.056$ & 0.100 & $-0.062$ & 0.102 &  & $-0.109$ & 0.108 & $-0.068$ & 0.103 & $-0.075$ & 0.105 \\
1.50 & $-0.111$ & 0.106 & $-0.072$ & 0.098 & $-0.078$ & 0.101 &  & $-0.123$ & 0.106 & $-0.084$ & 0.102 & $-0.090$ & 0.103 \\
1.54 & $-0.127$ & 0.105 & $-0.088$ & 0.097 & $-0.094$ & 0.099 &  & $-0.138$ & 0.104 & $-0.100$ & 0.100 & $-0.106$ & 0.102 \\
1.58 & $-0.143$ & 0.104 & $-0.105$ & 0.097 & $-0.110$ & 0.099 &  & $-0.153$ & 0.103 & $-0.115$ & 0.100 & $-0.121$ & 0.101 \\
1.62 & $-0.157$ & 0.103 & $-0.120$ & 0.097 & $-0.125$ & 0.099 &  & $-0.167$ & 0.103 & $-0.130$ & 0.099 & $-0.135$ & 0.101 \\
1.66 & $-0.170$ & 0.103 & $-0.133$ & 0.097 & $-0.139$ & 0.099 &  & $-0.179$ & 0.103 & $-0.142$ & 0.099 & $-0.148$ & 0.101 \\
1.70 & $-0.180$ & 0.104 & $-0.143$ & 0.098 & $-0.149$ & 0.100 &  & $-0.189$ & 0.104 & $-0.152$ & 0.100 & $-0.158$ & 0.102 \\
1.74 & $-0.187$ & 0.105 & $-0.150$ & 0.099 & $-0.156$ & 0.101 &  & $-0.195$ & 0.104 & $-0.158$ & 0.101 & $-0.164$ & 0.102 \\
1.78 & $-0.191$ & 0.105 & $-0.154$ & 0.100 & $-0.160$ & 0.101 &  & $-0.197$ & 0.104 & $-0.161$ & 0.101 & $-0.166$ & 0.102 \\
1.82 & $-0.196$ & 0.106 & $-0.159$ & 0.101 & $-0.165$ & 0.102 &  & $-0.200$ & 0.103 & $-0.163$ & 0.100 & $-0.169$ & 0.101 \\
1.87 & $-0.203$ & 0.108 & $-0.165$ & 0.103 & $-0.171$ & 0.104 &  & $-0.204$ & 0.103 & $-0.168$ & 0.100 & $-0.173$ & 0.102 \\
1.91 & $-0.207$ & 0.108 & $-0.169$ & 0.103 & $-0.175$ & 0.105 &  & $-0.209$ & 0.104 & $-0.172$ & 0.101 & $-0.177$ & 0.102 \\
1.96 & $-0.209$ & 0.107 & $-0.171$ & 0.102 & $-0.177$ & 0.104 &  & $-0.212$ & 0.105 & $-0.174$ & 0.102 & $-0.180$ & 0.103 \\
2.01 & $-0.211$ & 0.106 & $-0.173$ & 0.102 & $-0.179$ & 0.103 &  & $-0.214$ & 0.105 & $-0.177$ & 0.102 & $-0.183$ & 0.103 \\
2.06 & $-0.215$ & 0.106 & $-0.178$ & 0.102 & $-0.183$ & 0.103 &  & $-0.218$ & 0.104 & $-0.181$ & 0.101 & $-0.186$ & 0.103 \\
2.11 & $-0.220$ & 0.105 & $-0.184$ & 0.102 & $-0.189$ & 0.103 &  & $-0.222$ & 0.104 & $-0.186$ & 0.101 & $-0.191$ & 0.102 \\
2.16 & $-0.225$ & 0.105 & $-0.189$ & 0.102 & $-0.195$ & 0.103 &  & $-0.226$ & 0.103 & $-0.190$ & 0.100 & $-0.196$ & 0.101 \\
2.21 & $-0.226$ & 0.104 & $-0.191$ & 0.102 & $-0.196$ & 0.103 &  & $-0.226$ & 0.102 & $-0.191$ & 0.100 & $-0.197$ & 0.101 \\
2.26 & $-0.221$ & 0.105 & $-0.186$ & 0.102 & $-0.191$ & 0.104 &  & $-0.222$ & 0.103 & $-0.186$ & 0.100 & $-0.192$ & 0.101 \\
2.32 & $-0.211$ & 0.108 & $-0.174$ & 0.105 & $-0.180$ & 0.106 &  & $-0.211$ & 0.105 & $-0.173$ & 0.102 & $-0.179$ & 0.104 \\
2.38 & $-0.198$ & 0.109 & $-0.160$ & 0.106 & $-0.166$ & 0.108 &  & $-0.197$ & 0.107 & $-0.159$ & 0.103 & $-0.164$ & 0.105 \\
2.43 & $-0.187$ & 0.109 & $-0.148$ & 0.106 & $-0.154$ & 0.107 &  & $-0.185$ & 0.106 & $-0.146$ & 0.102 & $-0.152$ & 0.103 \\
2.49 & $-0.180$ & 0.108 & $-0.141$ & 0.105 & $-0.147$ & 0.107 &  & $-0.177$ & 0.105 & $-0.138$ & 0.100 & $-0.144$ & 0.102 \\
2.55 & $-0.187$ & 0.109 & $-0.148$ & 0.107 & $-0.154$ & 0.108 &  & $-0.183$ & 0.105 & $-0.143$ & 0.101 & $-0.150$ & 0.103 \\
2.61 & $-0.207$ & 0.112 & $-0.166$ & 0.110 & $-0.173$ & 0.111 &  & $-0.201$ & 0.107 & $-0.161$ & 0.103 & $-0.167$ & 0.104 \\
2.68 & $-0.224$ & 0.114 & $-0.183$ & 0.112 & $-0.189$ & 0.113 &  & $-0.218$ & 0.109 & $-0.177$ & 0.105 & $-0.183$ & 0.107 \\
2.74 & $-0.237$ & 0.116 & $-0.195$ & 0.114 & $-0.202$ & 0.115 &  & $-0.231$ & 0.112 & $-0.189$ & 0.107 & $-0.195$ & 0.109 \\
2.81 & $-0.245$ & 0.116 & $-0.204$ & 0.115 & $-0.210$ & 0.116 &  & $-0.239$ & 0.112 & $-0.196$ & 0.107 & $-0.203$ & 0.109 \\
2.88 & $-0.247$ & 0.114 & $-0.207$ & 0.113 & $-0.213$ & 0.114 &  & $-0.239$ & 0.109 & $-0.199$ & 0.105 & $-0.205$ & 0.106 \\
2.95 & $-0.244$ & 0.110 & $-0.208$ & 0.110 & $-0.213$ & 0.111 &  & $-0.236$ & 0.104 & $-0.197$ & 0.100 & $-0.203$ & 0.101 \\
3.02 & $-0.245$ & 0.107 & $-0.210$ & 0.109 & $-0.215$ & 0.109 &  & $-0.233$ & 0.099 & $-0.197$ & 0.096 & $-0.203$ & 0.098 \\
3.09 & $-0.254$ & 0.110 & $-0.221$ & 0.115 & $-0.226$ & 0.115 &  & $-0.235$ & 0.097 & $-0.201$ & 0.095 & $-0.206$ & 0.096 \\
3.17 & $-0.286$ & 0.129 & $-0.255$ & 0.142 & $-0.260$ & 0.141 &  & $-0.239$ & 0.097 & $-0.206$ & 0.095 & $-0.211$ & 0.096 \\
3.25 & $-0.341$ & 0.163 & $-0.310$ & 0.186 & $-0.315$ & 0.183 &  & $-0.246$ & 0.099 & $-0.213$ & 0.097 & $-0.218$ & 0.098 \\
3.32 & $-0.365$ & 0.175 & $-0.333$ & 0.201 & $-0.338$ & 0.197 &  & $-0.256$ & 0.102 & $-0.221$ & 0.100 & $-0.226$ & 0.101 \\
3.40 & $-0.341$ & 0.155 & $-0.308$ & 0.175 & $-0.313$ & 0.172 &  & $-0.266$ & 0.106 & $-0.230$ & 0.104 & $-0.235$ & 0.105 \\
3.49 & $-0.310$ & 0.132 & $-0.276$ & 0.143 & $-0.281$ & 0.142 &  & $-0.273$ & 0.109 & $-0.236$ & 0.107 & $-0.241$ & 0.108 \\
3.57 & $-0.298$ & 0.124 & $-0.264$ & 0.131 & $-0.269$ & 0.130 &  & $-0.275$ & 0.109 & $-0.238$ & 0.107 & $-0.244$ & 0.108 \\
3.66 & $-0.294$ & 0.121 & $-0.261$ & 0.128 & $-0.266$ & 0.127 &  & $-0.273$ & 0.107 & $-0.237$ & 0.105 & $-0.243$ & 0.106 \\
3.75 & $-0.291$ & 0.119 & $-0.260$ & 0.127 & $-0.265$ & 0.126 &  & $-0.269$ & 0.104 & $-0.235$ & 0.103 & $-0.240$ & 0.104 \\
3.84 & $-0.291$ & 0.119 & $-0.261$ & 0.128 & $-0.265$ & 0.127 &  & $-0.266$ & 0.102 & $-0.233$ & 0.101 & $-0.238$ & 0.102 \\
3.93 & $-0.297$ & 0.123 & $-0.267$ & 0.133 & $-0.271$ & 0.132 &  & $-0.268$ & 0.103 & $-0.235$ & 0.102 & $-0.240$ & 0.103 \\
4.03 & $-0.308$ & 0.129 & $-0.276$ & 0.140 & $-0.281$ & 0.139 &  & $-0.274$ & 0.106 & $-0.239$ & 0.104 & $-0.244$ & 0.105 \\
4.13 & $-0.316$ & 0.132 & $-0.283$ & 0.144 & $-0.288$ & 0.143 &  & $-0.280$ & 0.108 & $-0.244$ & 0.107 & $-0.250$ & 0.108 \\
4.23 & $-0.318$ & 0.134 & $-0.285$ & 0.147 & $-0.290$ & 0.146 &  & $-0.283$ & 0.111 & $-0.246$ & 0.109 & $-0.252$ & 0.110 \\
4.33 & $-0.316$ & 0.136 & $-0.284$ & 0.151 & $-0.289$ & 0.150 &  & $-0.279$ & 0.112 & $-0.241$ & 0.110 & $-0.247$ & 0.111 \\
4.43 & $-0.313$ & 0.139 & $-0.283$ & 0.157 & $-0.287$ & 0.154 &  & $-0.271$ & 0.111 & $-0.234$ & 0.110 & $-0.240$ & 0.111 \\
4.54 & $-0.312$ & 0.142 & $-0.284$ & 0.162 & $-0.288$ & 0.159 &  & $-0.265$ & 0.110 & $-0.229$ & 0.108 & $-0.234$ & 0.109 \\
4.65 & $-0.316$ & 0.145 & $-0.289$ & 0.166 & $-0.293$ & 0.164 &  & $-0.263$ & 0.108 & $-0.228$ & 0.107 & $-0.233$ & 0.108 \\
4.77 & $-0.323$ & 0.148 & $-0.297$ & 0.172 & $-0.301$ & 0.169 &  & $-0.264$ & 0.107 & $-0.229$ & 0.106 & $-0.235$ & 0.107 \\
4.88 & $-0.333$ & 0.154 & $-0.307$ & 0.179 & $-0.311$ & 0.176 &  & $-0.265$ & 0.107 & $-0.231$ & 0.106 & $-0.236$ & 0.107 \\
5.00 & $-0.347$ & 0.162 & $-0.323$ & 0.190 & $-0.326$ & 0.187 &  & $-0.267$ & 0.106 & $-0.233$ & 0.106 & $-0.238$ & 0.106  
\enddata
\tablecomments{
Col. (1): Central wavelength of the luminosity in units of $\mu$m.
Col. (2): logarithmic $M_*/L$ of the DustPedia sample based on the total luminosity.
Col. (3): Scatter (dex) in $M_*/L$ of the DustPedia sample based on the total luminosity.
Col. (4): logarithmic $M_*/L$ of the S82 sample based on the total luminosity.
Col. (5): Scatter (dex) in logarithmic $M_*/L$ of the S82 sample based on the total luminosity.
Col. (6): logarithmic $M_*/L$ of the entire sample based on the total luminosity.
Col. (7): Scatter (dex) in logarithmic $M_*/L$ of the entire sample based on the total luminosity.
Col. (8): $M_*/L$ of the DustPedia sample based on the stellar luminosity.
Col. (9): Scatter (dex) in logarithmic $M_*/L$ of the DustPedia sample based on the stellar luminosity.
Col. (10): $M_*/L$ of the S82 sample based on the stellar luminosity.
Col. (11): Scatter (dex) in logarithmic $M_*/L$ of the S82 sample based on the stellar luminosity.
Col. (12): $M_*/L$ of the entire sample based on the stellar luminosity.
Col. (13): Scatter (dex) in logarithmic $M_*/L$ of the entire sample based on the stellar luminosity.
}
\end{deluxetable*}

\clearpage

\startlongtable
\begin{deluxetable}{cccc}
\scriptsize
\tablecaption{$\log\ (M_*/L)=a+b\log\ (M_*/M_\odot)$\label{tab:table2}}
\tablehead{
$\lambda_c$ & $a$ & $b$ & $\sigma$ \\
(1) & (2) & (3) & (4) 
}
\startdata
0.75 & $-1.344$ & 0.141 & 0.130 \\
0.77 & $-1.338$ & 0.139 & 0.129 \\
0.79 & $-1.327$ & 0.138 & 0.127 \\
0.81 & $-1.297$ & 0.134 & 0.123 \\
0.83 & $-1.277$ & 0.130 & 0.121 \\
0.85 & $-1.270$ & 0.129 & 0.119 \\
0.87 & $-1.251$ & 0.126 & 0.117 \\
0.89 & $-1.219$ & 0.122 & 0.114 \\
0.91 & $-1.184$ & 0.118 & 0.111 \\
0.93 & $-1.156$ & 0.114 & 0.109 \\
0.95 & $-1.136$ & 0.112 & 0.107 \\
0.98 & $-1.116$ & 0.109 & 0.106 \\
1.00 & $-1.090$ & 0.106 & 0.104 \\
1.02 & $-1.065$ & 0.103 & 0.103 \\
1.05 & $-1.052$ & 0.101 & 0.102 \\
1.08 & $-1.046$ & 0.100 & 0.101 \\
1.10 & $-1.040$ & 0.099 & 0.100 \\
1.13 & $-1.036$ & 0.099 & 0.100 \\
1.16 & $-1.030$ & 0.098 & 0.100 \\
1.18 & $-1.021$ & 0.096 & 0.099 \\
1.21 & $-1.010$ & 0.095 & 0.098 \\
1.24 & $-1.003$ & 0.094 & 0.098 \\
1.27 & $-1.001$ & 0.093 & 0.097 \\
1.30 & $-1.005$ & 0.093 & 0.097 \\
1.33 & $-1.012$ & 0.093 & 0.097 \\
1.37 & $-1.019$ & 0.093 & 0.097 \\
1.40 & $-1.024$ & 0.093 & 0.096 \\
1.43 & $-1.019$ & 0.092 & 0.095 \\
1.47 & $-1.008$ & 0.089 & 0.094 \\
1.50 & $-0.993$ & 0.087 & 0.093 \\
1.54 & $-0.982$ & 0.084 & 0.092 \\
1.58 & $-0.979$ & 0.082 & 0.091 \\
1.62 & $-0.974$ & 0.080 & 0.091 \\
1.66 & $-0.966$ & 0.078 & 0.092 \\
1.70 & $-0.965$ & 0.077 & 0.093 \\
1.74 & $-0.971$ & 0.077 & 0.094 \\
1.78 & $-0.973$ & 0.077 & 0.094 \\
1.82 & $-0.974$ & 0.077 & 0.093 \\
1.87 & $-0.984$ & 0.078 & 0.093 \\
1.91 & $-0.997$ & 0.079 & 0.094 \\
1.96 & $-1.005$ & 0.079 & 0.094 \\
2.01 & $-1.004$ & 0.079 & 0.094 \\
2.06 & $-0.993$ & 0.077 & 0.094 \\
2.11 & $-0.979$ & 0.076 & 0.094 \\
2.16 & $-0.961$ & 0.073 & 0.094 \\
2.21 & $-0.947$ & 0.072 & 0.093 \\
2.26 & $-0.958$ & 0.073 & 0.094 \\
2.32 & $-0.993$ & 0.078 & 0.095 \\
2.38 & $-1.023$ & 0.082 & 0.095 \\
2.43 & $-1.034$ & 0.085 & 0.093 \\
2.49 & $-1.050$ & 0.087 & 0.091 \\
2.55 & $-1.076$ & 0.089 & 0.092 \\
2.61 & $-1.106$ & 0.090 & 0.093 \\
2.68 & $-1.138$ & 0.092 & 0.095 \\
2.74 & $-1.163$ & 0.093 & 0.097 \\
2.81 & $-1.165$ & 0.092 & 0.098 \\
2.88 & $-1.131$ & 0.089 & 0.096 \\
2.95 & $-1.071$ & 0.083 & 0.092 \\
3.02 & $-1.013$ & 0.078 & 0.089 \\
3.09 & $-0.977$ & 0.074 & 0.088 \\
3.17 & $-0.958$ & 0.072 & 0.089 \\
3.25 & $-0.955$ & 0.071 & 0.091 \\
3.32 & $-0.975$ & 0.072 & 0.094 \\
3.40 & $-1.007$ & 0.074 & 0.098 \\
3.49 & $-1.025$ & 0.075 & 0.100 \\
3.57 & $-1.019$ & 0.074 & 0.101 \\
3.66 & $-0.992$ & 0.072 & 0.099 \\
3.75 & $-0.955$ & 0.069 & 0.097 \\
3.84 & $-0.932$ & 0.067 & 0.096 \\
3.93 & $-0.938$ & 0.067 & 0.097 \\
4.03 & $-0.966$ & 0.069 & 0.099 \\
4.13 & $-0.999$ & 0.072 & 0.101 \\
4.23 & $-1.020$ & 0.074 & 0.103 \\
4.33 & $-1.019$ & 0.074 & 0.104 \\
4.43 & $-0.999$ & 0.073 & 0.104 \\
4.54 & $-0.972$ & 0.071 & 0.103 \\
4.65 & $-0.951$ & 0.069 & 0.102 \\
4.77 & $-0.942$ & 0.068 & 0.101 \\
4.88 & $-0.935$ & 0.067 & 0.101 \\
5.00 & $-0.929$ & 0.066 & 0.101
\enddata
\tablecomments{
Col. (1): Central wavelength of the luminosity in units of $\mu$m.
Col. (2): Interceptor in the linear regression fit between logarithmic stellar mass in units of the solar mass and logarithmic $M_*/L$. 
Col. (3): Slope in the linear regression fit between logarithmic stellar mass in units of the solar mass and logarithmic $M_*/L$. 
Col. (4): Scatter (dex) in logarithmic $M_*/L$.
}
\end{deluxetable}

\clearpage

\startlongtable
\begin{deluxetable}{cccccc}
\scriptsize
\tablecaption{$\log\ (M_*/L)= a \left\{\frac{1}{2}\left[1+\left( \frac{\log\ {\rm sSFR}}{\delta}\right)^{1/\beta}\right] \right\}^{-\alpha\beta}$\label{tab:table3}}
\tablehead{
$\lambda_c$ & $a$ & $\delta$ & $\beta$ & $\alpha$ & $\sigma$ \\
(1) & (2) & (3) & (4) & (5) & (6) }
\startdata
0.75 & $0.180$ & 0.0114 & 1.041 & 0.320 & 0.057 \\
0.77 & $0.172$ & 0.0113 & 1.053 & 0.315 & 0.058 \\
0.79 & $0.163$ & 0.0113 & 1.064 & 0.309 & 0.059 \\
0.81 & $0.151$ & 0.0109 & 1.046 & 0.298 & 0.056 \\
0.83 & $0.141$ & 0.0106 & 1.031 & 0.290 & 0.054 \\
0.85 & $0.132$ & 0.0104 & 1.027 & 0.285 & 0.053 \\
0.87 & $0.123$ & 0.0101 & 0.999 & 0.279 & 0.049 \\
0.89 & $0.114$ & 0.0097 & 0.953 & 0.270 & 0.044 \\
0.91 & $0.106$ & 0.0093 & 0.911 & 0.262 & 0.038 \\
0.93 & $0.101$ & 0.0090 & 0.879 & 0.255 & 0.035 \\
0.95 & $0.095$ & 0.0087 & 0.852 & 0.249 & 0.032 \\
0.98 & $0.090$ & 0.0084 & 0.824 & 0.244 & 0.029 \\
1.00 & $0.085$ & 0.0082 & 0.795 & 0.238 & 0.027 \\
1.02 & $0.080$ & 0.0080 & 0.768 & 0.233 & 0.025 \\
1.05 & $0.075$ & 0.0079 & 0.752 & 0.229 & 0.024 \\
1.08 & $0.071$ & 0.0077 & 0.746 & 0.226 & 0.024 \\
1.10 & $0.069$ & 0.0076 & 0.742 & 0.223 & 0.023 \\
1.13 & $0.066$ & 0.0075 & 0.736 & 0.221 & 0.023 \\
1.16 & $0.062$ & 0.0074 & 0.726 & 0.219 & 0.023 \\
1.18 & $0.058$ & 0.0073 & 0.717 & 0.217 & 0.022 \\
1.21 & $0.054$ & 0.0072 & 0.708 & 0.214 & 0.022 \\
1.24 & $0.049$ & 0.0071 & 0.698 & 0.211 & 0.022 \\
1.27 & $0.044$ & 0.0070 & 0.689 & 0.209 & 0.022 \\
1.30 & $0.038$ & 0.0070 & 0.682 & 0.208 & 0.022 \\
1.33 & $0.032$ & 0.0069 & 0.681 & 0.208 & 0.022 \\
1.37 & $0.026$ & 0.0068 & 0.684 & 0.207 & 0.022 \\
1.40 & $0.020$ & 0.0068 & 0.691 & 0.205 & 0.022 \\
1.43 & $0.010$ & 0.0067 & 0.692 & 0.202 & 0.022 \\
1.47 & $-0.002$ & 0.0066 & 0.680 & 0.198 & 0.022 \\
1.50 & $-0.017$ & 0.0065 & 0.658 & 0.193 & 0.022 \\
1.54 & $-0.033$ & 0.0064 & 0.639 & 0.189 & 0.022 \\
1.58 & $-0.049$ & 0.0064 & 0.624 & 0.186 & 0.023 \\
1.62 & $-0.062$ & 0.0063 & 0.605 & 0.184 & 0.024 \\
1.66 & $-0.074$ & 0.0063 & 0.582 & 0.183 & 0.027 \\
1.70 & $-0.083$ & 0.0063 & 0.564 & 0.183 & 0.029 \\
1.74 & $-0.089$ & 0.0063 & 0.559 & 0.184 & 0.030 \\
1.78 & $-0.092$ & 0.0062 & 0.565 & 0.183 & 0.030 \\
1.82 & $-0.095$ & 0.0062 & 0.574 & 0.182 & 0.029 \\
1.87 & $-0.099$ & 0.0062 & 0.578 & 0.183 & 0.028 \\
1.91 & $-0.103$ & 0.0062 & 0.577 & 0.184 & 0.029 \\
1.96 & $-0.105$ & 0.0062 & 0.577 & 0.185 & 0.029 \\
2.01 & $-0.107$ & 0.0062 & 0.573 & 0.185 & 0.029 \\
2.06 & $-0.111$ & 0.0062 & 0.563 & 0.183 & 0.030 \\
2.11 & $-0.117$ & 0.0062 & 0.550 & 0.181 & 0.032 \\
2.16 & $-0.122$ & 0.0062 & 0.538 & 0.179 & 0.034 \\
2.21 & $-0.123$ & 0.0062 & 0.532 & 0.177 & 0.034 \\
2.26 & $-0.117$ & 0.0062 & 0.537 & 0.179 & 0.034 \\
2.32 & $-0.103$ & 0.0062 & 0.555 & 0.185 & 0.031 \\
2.38 & $-0.088$ & 0.0062 & 0.581 & 0.189 & 0.028 \\
2.43 & $-0.077$ & 0.0063 & 0.621 & 0.189 & 0.024 \\
2.49 & $-0.072$ & 0.0063 & 0.677 & 0.189 & 0.022 \\
2.55 & $-0.079$ & 0.0064 & 0.710 & 0.190 & 0.022 \\
2.61 & $-0.094$ & 0.0063 & 0.703 & 0.193 & 0.023 \\
2.68 & $-0.108$ & 0.0062 & 0.685 & 0.196 & 0.023 \\
2.74 & $-0.118$ & 0.0062 & 0.672 & 0.200 & 0.024 \\
2.81 & $-0.125$ & 0.0062 & 0.663 & 0.200 & 0.024 \\
2.88 & $-0.129$ & 0.0062 & 0.657 & 0.194 & 0.024 \\
2.95 & $-0.131$ & 0.0061 & 0.651 & 0.185 & 0.023 \\
3.02 & $-0.133$ & 0.0061 & 0.636 & 0.177 & 0.024 \\
3.09 & $-0.136$ & 0.0061 & 0.610 & 0.172 & 0.026 \\
3.17 & $-0.141$ & 0.0060 & 0.581 & 0.170 & 0.029 \\
3.25 & $-0.146$ & 0.0060 & 0.555 & 0.171 & 0.032 \\
3.32 & $-0.152$ & 0.0061 & 0.535 & 0.176 & 0.035 \\
3.40 & $-0.063$ & 0.0155 & 0.007 & 0.259 & 0.044 \\
3.49 & $-0.062$ & 0.0153 & 0.007 & 0.259 & 0.046 \\
3.57 & $-0.062$ & 0.0153 & 0.007 & 0.258 & 0.047 \\
3.66 & $-0.062$ & 0.0155 & 0.007 & 0.258 & 0.047 \\
3.75 & $-0.062$ & 0.0158 & 0.007 & 0.257 & 0.048 \\
3.84 & $-0.062$ & 0.0159 & 0.007 & 0.255 & 0.048 \\
3.93 & $-0.062$ & 0.0159 & 0.007 & 0.256 & 0.048 \\
4.03 & $-0.062$ & 0.0157 & 0.007 & 0.256 & 0.049 \\
4.13 & $-0.062$ & 0.0155 & 0.007 & 0.257 & 0.049 \\
4.23 & $-0.062$ & 0.0152 & 0.007 & 0.257 & 0.050 \\
4.33 & $-0.062$ & 0.0151 & 0.007 & 0.258 & 0.051 \\
4.43 & $-0.159$ & 0.0060 & 0.487 & 0.185 & 0.047 \\
4.54 & $-0.155$ & 0.0060 & 0.482 & 0.182 & 0.048 \\
4.65 & $-0.155$ & 0.0060 & 0.477 & 0.179 & 0.048 \\
4.77 & $-0.157$ & 0.0060 & 0.474 & 0.177 & 0.048 \\
4.88 & $-0.062$ & 0.0156 & 0.007 & 0.257 & 0.053 \\
5.00 & $-0.062$ & 0.0157 & 0.007 & 0.257 & 0.053
\enddata
\tablecomments{
Col. (1): Central wavelength of the luminosity in units of $\mu$m.
Col. (2): Normalization factor in the smoothly broken power law fit, which is equivalent to $\log\ (M_*/L)$ at ${\rm sSFR} \ll \delta$. Note that the unit of sSFR is Gyr$^{-1}$.
Col. (3): Pivot sSFR in units of Gyr$^{-1}$.
Col. (4): Smoothness parameter.
Col. (5): Power-law index above the pivot sSFR.
Col. (6): Scatter (dex) in logarithmic $M_*/L$.
}
\end{deluxetable}

\end{document}